\documentclass[prd,aps,showpacs,twocolumn]{revtex4-1}
\usepackage{bm}
\usepackage{amsmath}
\usepackage{amssymb}
\usepackage{graphicx}
\usepackage{slashed}
\usepackage{accents}
\usepackage{mathrsfs}
\usepackage{braket}
\usepackage{units}
\usepackage{upgreek}
\usepackage[caption=false]{subfig}

\usepackage[breaklinks=true]{hyperref}
\usepackage{breakurl} 

\DeclareSymbolFont{eulerscript}{U}{eur}{m}{n}
\DeclareSymbolFontAlphabet{\matheuler}{eulerscript}
\DeclareMathAlphabet{\boldgreek}{OML}{zplm}{b}{it}

\newcommand{\Q}{\mathcal{Q}}
\newcommand{\I}{i}
\newcommand{\nfrac}[2]{{#1}/{#2}}
\newcommand{\one}{\mathbf{1}}
\newcommand{\del}{\partial}
\newcommand{\eps}{\epsilon}
\newcommand{\mc}[1]{\mathcal{#1}}
\newcommand{\spvec}[1]{\boldsymbol{#1}}
\newcommand{\lb}{\left(}
\newcommand{\rb}{\right)}
\newcommand{\rsb}{\right]}
\newcommand{\lsb}{\left[}
\newcommand{\abs}[1]{\left|#1\right|}
\DeclareMathOperator{\tr}{\mathbf{tr}}
\DeclareMathOperator{\Hankel}{H}
\DeclareMathOperator{\PP}{P}
\DeclareMathOperator{\sign}{sign}
\DeclareMathOperator{\Ai}{Ai}
\DeclareMathOperator{\Gi}{Gi}
\newcommand{\probsym}{W}
\newcommand{\lplus}{{{}+}}
\newcommand{\lminus}{{{}-}}
\newcommand{\lone}{{\scalebox{.64}{$\matheuler{I}$}}}
\newcommand{\ltwo}{{\scalebox{.64}{$\matheuler{II}$}}}
\newcommand{\lperp}{\perp}
\newcommand{\s}[1]{\slashed{#1}}
\newcommand{\Ftilde}{\mathfrak{F}}
\newcommand{\highlight}[1]{#1}

\begin{document}
\title{Polarization-operator approach to pair creation in short laser pulses}
\author{Sebastian \surname{Meuren}}
\email{s.meuren@mpi-hd.mpg.de}
\author{Karen Z. \surname{Hatsagortsyan}}
\email{k.hatsagortsyan@mpi-hd.mpg.de}
\author{Christoph H. \surname{Keitel}}
\email{keitel@mpi-hd.mpg.de}
\author{Antonino \surname{Di Piazza}}
\email{dipiazza@mpi-hd.mpg.de}
\affiliation{Max-Planck-Institut f\"ur Kernphysik, Saupfercheckweg 1, D-69117 Heidelberg, Germany}
\date{\today}

\begin{abstract}
Short-pulse effects are investigated for the nonlinear Breit-Wheeler process, i.e. the production of an electron-positron pair induced by a gamma photon inside an intense plane-wave laser pulse. To obtain the total pair-creation probability we verify (to leading-order) the cutting rule for the polarization operator in the realm of strong-field QED by an explicit calculation. Using a double-integral representation for the leading-order contribution to the polarization operator, compact expressions for the total pair-creation probability inside an arbitrary plane-wave background field are derived. \highlight{Correspondingly, the photon wave function including leading-order radiative corrections in the laser field is obtained via the Schwinger-Dyson equation in the quasistatic approximation. Moreover, the influence of the carrier-envelope phase (CEP) and of the laser pulse shape on the total pair-creation probability in a linearly polarized laser pulse is investigated, and the validity of the (local) constant-crossed field approximation analyzed. It is shown that with presently available technology pair-creation probabilities of the order of ten percent could be reached for a single gamma photon.}
 
\end{abstract}

\maketitle

\section{Introduction}

\begin{figure}[b]
\centering
\includegraphics{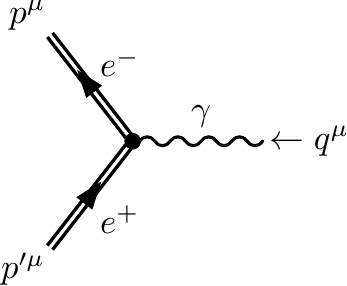}
\caption{\label{fig:pair}Leading-order Feynman diagram for the creation of an electron-positron pair by a single photon inside a plane-wave background field. The fermions are denoted by double lines, representing the Volkov states (solutions of the Dirac equation, which take the plane-wave background field into account exactly) \cite{volkov_ueber_1935}. The four-vectors indicate the four-momenta of the particles; they are described in Sec. \ref{sec:paircreationsection}.}
\end{figure}

In vacuum the decay of a single photon into a real electron-positron pair  is forbidden by energy-momentum conservation, even if the energy $\hbar\omega_\gamma$ of the photon exceeds the threshold $2mc^2$ ($m$ denotes the electron mass). At least one additional interaction is needed to catalyze the process, e.g. a second photon (Breit-Wheeler process \cite{breit_collision_1934}) or the Coulomb field of a nucleus (Bethe-Heitler pair creation \cite{oppenheimer_production_1933,bethe_stopping_1934}). 

Inside a strong electromagnetic background field the situation changes, as the field can provide four-momentum to the reaction. If a constant electric field reaches the critical field strength $E_{cr}=\nfrac{m^2c^3}{(\hbar|e|)} = \unitfrac[1.3\times 10^{16}]{V}{cm}$ of QED, even the spontaneous creation of electron-positron pairs from the vacuum  becomes possible ($e<0$ denotes the electron charge) \cite{sauter_ueber_1931,heisenberg_folgerungen_1936,schwinger_gauge_1951,di_piazza_extremely_2012}. Experimentally, spontaneous pair creation could be observed if an intense x-ray laser is focused to its diffraction limit \cite{alkofer_pair_2001,ringwald_pair_2001}. 

However, present and near-future laser facilities will not reach the critical field strength ($\unitfrac[10^{24}-10^{25}]{W}{cm^2}$ is the envisaged intensity for optical lasers \cite{ELI,CLF,XCELS} and the critical field corresponds to an intensity of~$I_{\mathrm{cr}}=\eps_0 c E_{\mathrm{cr}}^2 =\unitfrac[4.6\times 10^{29}]{W}{cm^2}$). Nevertheless, electron-positron pairs could be produced if the process is stimulated by a highly energetic particle. For example, in the E-144 experiment at SLAC electron-positron pair creation has been observed during the collision of an electron beam with a relativistically intense optical laser via the trident process \cite{burke_positron_1997,hu_complete_2010,ilderton_trident_2011,king_trident_2013}.

In this paper we consider the nonlinear Breit-Wheeler process shown in Fig.~\ref{fig:pair}, i.e. pair creation by a single (on shell) gamma photon inside a strong (optical) laser pulse with electric field amplitude $E_0$ and central angular frequency $\omega$. \highlight{By absorbing multiple low-energy laser photons, the decay of the photon into an electron-positron pair, which is forbidden in vacuum, becomes feasible.} For monochromatic laser fields this process has been considered in detail (see e.g. \cite{reiss_absorption_1962,nikishov_quantum_1964,nikishov_quantum2_1964,narozhny_quantum_1965,nikishov_pair_1967,ritus_1985}). In the strong-field regime $\xi \gg 1$, where in general many photons can be efficiently absorbed from the laser field,  the probability depends nontrivially only on the quantum-nonlinearity parameter $\chi$ and is exponentially suppressed for $\chi \ll 1$. Here, $\xi=\nfrac{|e|E_0}{(m\omega c)}$ is a gauge and Lorentz invariant measure of the laser intensity \cite{heinzl_lorentz_2009} and $\chi=(2\nfrac{\hbar\omega_\gamma}{mc^2}) (\nfrac{E_0}{E_{cr}})$ for the head-on collision between a gamma photon with energy $\hbar\omega_\gamma$ and the laser pulse. As existing optical petawatt laser systems reach already $\xi\sim 100$ \cite{yanovsky_ultra_2008} and $\unit{GeV}$ photons are available using Compton backscattering \cite{leemans_gev_2006,esarey_physics_2009,phuoc_all-optical_2012,wang_quasi-monoenergetic_2013,powers_quasi-monoenergetic_2014,muramatsu_development_2014}, the regime $\chi \gtrsim 1$ could be entered with available technology. 

Due to the experimental progress concerning laser development during the last years, the nonlinear Breit-Wheeler process has been recently investigated by several authors \cite{schutzhold_dynamically_2008,dunne_catalysis_2009,bulanov_multiple_2010,heinzl_finite_2010,tuchin_non-linear_2010,orthaber_momentum_2011,ipp_streaking_2011,hebenstreit_pair_2011,krajewska_breit-wheeler_2012,titov_enhanced_2012,nousch_pair_2012,king_photon_2013,bulanov_electromagnetic_2013,jansen_strongly_2013,fedotov_pair_2013,titov_breit-wheeler_2013,kohlfurst_effective_2014,krajewska_breit-wheeler_2014} (see also the reviews \cite{di_piazza_extremely_2012,ruffini_electronpositron_2010}). To achieve the strong field strengths needed to observe the nonlinear Breit-Wheeler process, future experiments will probably use short laser pulses. However, the calculation of the total pair-creation probability is challenging if the phase-space integrals are calculated numerically for an arbitrary plane-wave field \cite{titov_enhanced_2012}. 

In the present paper we circumvent these difficulties by applying the optical theorem to the polarization operator (see Fig. \ref{fig:polop}) \cite{baier_interaction_1975,becker_vacuum_1975,meuren_polarization_2013,dinu_vacuum_2014,dinu_photon_2014,gies_laser_2014}. The optical theorem, which relates the total probability for particle production processes to the imaginary part of corresponding loop diagrams, reflects the unitarity of the $S$-matrix. \highlight{As probability is conserved, the total pair-creation probability must be related to the imaginary part of the forward-scattering amplitude for photons (i.e. the polarization operator) \cite{ritus_1985,fradkin_quantum_1991}. However, it is instructive to verify this by an explicit calculation, which leads to the so-called ``cutting'' or ``cutkosky rules'' (for QED in vacuum this derivation was first given in \cite{landau_analytic_1959,cutkosky_singularities_1960}).}

By applying the cutting rule to the polarization operator we derive a double-integral representation for the total pair-creation probability inside a plane-wave laser pulse (for pair creation in combined laser and Coulomb fields the same method was used in \cite{di_piazza_barrier_2009,milstein_polarization-operator_2006}, see also \cite{baier_interaction_1975,dinu_vacuum_2014}). The analysis holds for an arbitrarily shaped plane-wave background field, in particular we focus on the description of experiments with short (optical) laser pulses. \highlight{Similar methods for  calculating analytically the final-state momentum integrals have recently also been applied in \cite{dinu_exact_2013,dinu_vacuum_2014}.}

\highlight{Starting from this compact representation of the total pair-creation probability we investigate various parameter regimes numerically. It is well known that for $\xi \gg 1$ the total pair-creation probability can be obtained by averaging over the corresponding result in a constant-crossed field \cite{ritus_1985,di_piazza_extremely_2012}. By comparing with the exact expression we show that already for $\xi \gtrsim 1$ this (local) constant-crossed field approximation may be applied. However, the importance of CEP effects is underestimated for $\xi \lesssim 1$ (in this regime the formation region for the pair-production process becomes large, which is not included in the constant-crossed field limit). Furthermore, the influence of the pulse shape, length and CEP on the total pair-creation probability are studied in the regime $\xi \gg 1$.}

\highlight{If the total pair-production probability becomes of order unity, the straight-forward evaluation of the diagram shown in Fig. \ref{fig:pair} is not sufficient. In this regime one must take into account that the exact photon wave function decays exponentially if pair-creation is possible. This exponential decay is naturally obtained by solving the Schwinger-Dyson equation for the exact photon wave function \cite{schwinger_gauge_1951,landau_quantum_1981}. We show that already for available laser parameters and photon energies this effect plays an important role (see also \cite{dinu_vacuum_2014}).}

The paper is organized as follows. The optical theorem is discussed in Sec. \ref{sec:paircreation_opticaltheorem} (see also the Appendices \ref{sec:paircreationappendix} and \ref{sec:cuttingrulesappendix}). \highlight{In Sec. \ref{sec:exactphotonwavefunction} the exponential decay of the exact photon wave function is derived.} In Sec. \ref{sec:paircreationprobability} the optical theorem is used to obtain a double-integral representation for the total pair-creation probability. Finally, numerical results are presented in Sec. \ref{sec:numericalresults}. Further details are given in the Appendices, in particular the double-integral representation for the polarization operator is discussed in Appendix \ref{sec:poloppaperresult} \highlight{(for other double-integral representations see \cite{dinu_vacuum_2014,becker_vacuum_1975})}.

From now on we use natural units $\hbar = c = 1$ and Heaviside-Lorentz units for charge [$\alpha = \nfrac{e^2}{(4\pi)} \approx \nfrac{1}{137}$ denotes the fine-structure constant], the notation agrees with \cite{meuren_polarization_2013}.

\section{Optical theorem}
\label{sec:paircreation_opticaltheorem}

\subsection{Pair creation with background fields}
\label{sec:paircreationsection}

The leading-order Feynman diagram for the creation of an electron and a positron with four-momenta $p^\mu$ and $p'^\mu$, respectively, by a photon with four-momentum $q^\mu$ is shown in Fig.~\ref{fig:pair}. In vacuum this process is forbidden, as four-momentum conservation $p^\mu + p'^\mu = q^\mu$ cannot be fulfilled if all three particles are on shell (i.e. $p^2=p'^2=m^2$, $q^2=0$). However, inside a plane-wave laser field additional laser photons with average four-momentum $k^\mu$ can be absorbed [see Eq.~(\ref{eqn:dressedvertexfinal_momentumconservation})]
\begin{gather}
p^\mu + p'^\mu = q^\mu + n k^\mu
\end{gather}
and the process of stimulated pair creation by an incoming photon is feasible (as a non-monochromatic plane-wave laser pulse consists of photons with different four-momenta, $n$ must not be an integer in general). 

\highlight{As shown in appendix \ref{sec:paircreationappendix}, the total pair-creation probability is given by [see Eq. (\ref{eqn:paircreation_decayratefinalA})]
\begin{subequations}
\label{eqn:paircreation_decayratefinalAcopy}
\begin{gather}
\probsym
=
\int \frac{d^3q'}{(2\pi)^3\, 2\eps_{\spvec{q}'}} \abs{\eta(q')}^2 \, W(q'),
\end{gather}
where
\begin{multline}
\probsym(q)
=
\sum_{\text{spin}} \int \frac{d^3p \, d^3p'}{(2\pi)^6 \, 2\eps_{\spvec{p}} 2\eps_{\spvec{p}'}} \,
\frac{1}{2q^\lminus} \, \abs{\mc{M}(p,p';q)}^2 
\\ \times \,
(2\pi)^3 \delta^{(\lminus,\lperp)}(p+p'-q)
\end{multline}
\end{subequations}
and $\eta(q')$ describes the (normalized) momentum distribution of the photon wave packet [see Eq. (\ref{eqn:paircreation_photonwavepacketstate}) and Eq. (\ref{eqn:paircreation_wavepacketenvelopenormalization}) for details]. We point out that the interpretation of $\probsym$ changes if it becomes of order unity, see Sec. \ref{sec:exactphotonwavefunction}.}

The reduced matrix element reads (to leading order, see Fig.~\ref{fig:pair})
\begin{gather}
\label{eqn:paircreation_reducedmatrixelementleadingorder}
\mc{M}(p,p';q) = \eps_\mu \, \bar{u}_{p} \mc{G}^\mu(p,q,-p') v_{p'}
\end{gather}
where $\eps^\mu$ denotes the polarization four-vector of the incoming photon, $u_p$ and $v_{p'}$ the Dirac spinors of the electron and the positron, respectively and $\mc{G}^\mu$ the nonsingular part of the dressed vertex [see Eqs. (\ref{eqn:dressedvertexfinal_momentumconservation}), (\ref{eqn:paircreation_reducedmatrixelement}) and (\ref{eqn:paircreation_matrixelement})].

Thus, $\probsym$ is the average of $\probsym(q)$ over the momentum distribution of the incoming photon wave packet. Assuming that the matrix element is sufficiently smooth and that the wave packet is peaked around $q'^\mu = q^\mu$, we obtain $W \approx W(q)$ [see Eq.~(\ref{eqn:paircreation_wavepacketenvelopenormalization})]. We point out that the average over the momentum distribution of the incoming gamma photon may hide substructures in the spectrum, especially if the energy spread of the incoming gamma photon ($\sim \unit{MeV}$) is much larger than the energy of the colliding laser photons ($\sim \unit{eV}$).

\subsection{Cutting rule}

\begin{figure}
\centering
\includegraphics{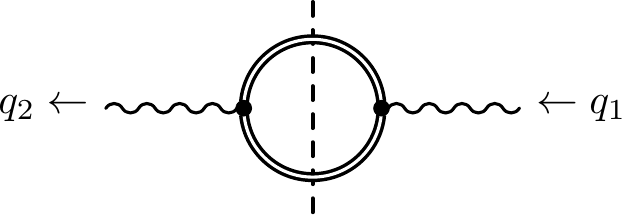}
\caption{\label{fig:polop}Leading-order contribution to the polarization operator for a photon in a plane-wave background field. The double lines denote the dressed (Volkov) propagator, which takes the classical background field into account exactly. The imaginary part of this diagram is related to the total pair-creation probability via the optical theorem [indicated by the dashed line, see Eq.~(\ref{eqn:paircreation_opticaltheorem})].}
\end{figure}

We obtain the differential pair-creation probability by inserting the pair-creation matrix element given in Eq.~(\ref{eqn:paircreation_reducedmatrixelementleadingorder}) into Eq.~(\ref{eqn:paircreation_decayratefinalAcopy}). To determine the total pair-creation probability, the phase-space integrals must be evaluated, which is numerically rather demanding \cite{titov_enhanced_2012}. If one is only interested in the total pair-creation probability (and not in its differential structure), these integrals can be taken analytically by applying cutting rules to the polarization operator, \highlight{as shown in appendix \ref{sec:cuttingrulesappendix} (for the application of cutting rules see also \cite{ritus_vacuum_1972,baier_interaction_1975,fradkin_quantum_1991,di_piazza_barrier_2009,milstein_polarization-operator_2006,dinu_vacuum_2014} and \cite{dinu_exact_2013} for an alternative method to solve the final momentum integrals analytically).}

Finally,  we obtain the following relation between the total nonlinear Breit-Wheeler pair-creation probability $W$ [see Eq. (\ref{eqn:paircreation_decayratefinalAcopy})] and the imaginary part of the photon forward-scattering amplitude [see Eq. (\ref{eqn:paircreation_opticaltheorem}) and also \cite{di_piazza_barrier_2009,milstein_polarization-operator_2006,baier_interaction_1975}]
\begin{gather}
\label{eqn:paircreation_opticaltheoremmaintext}
\probsym(q)
=
\frac{1}{kq} \, \Im \, [\eps_\mu \eps^*_\nu \, \Pi^{\mu\nu}(q,q)]
\end{gather}
($q^2=0$, $kq > 0$).

\subsection{Generalization}

\begin{figure}
\centering
\includegraphics{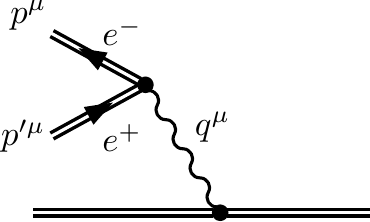}
\caption{\label{fig:trident}Leading-order Feynman diagram for the creation of an electron-positron pair by an electron or a positron inside a background field (trident process). The intermediate photon is in general not on shell (time axis from right to left).}
\end{figure}

The optical theorem given in Eq.~(\ref{eqn:paircreation_opticaltheoremmaintext}) holds for the decay of a real photon into a lepton pair. We will now generalize this result to the case where the pair-creation process represents only a part of a more complicated Feynman diagram. The most prominent example is the trident process shown in Fig. \ref{fig:trident} \cite{hu_complete_2010,ilderton_trident_2011,king_trident_2013}. After squaring the matrix element and summing over the final spin states, one has to consider the expression [see Eqs.~(\ref{eqn:paircreation_matrixelementsquared}) and (\ref{eqn:paircreation_imforwardscatteringamplitude})]
\begin{multline}
\mathfrak{M}^{\mu\nu}(q_1,q_2) = \int \frac{d^3p \, d^3p'}{(2\pi)^6 \, 2\eps_{\spvec{p}} 2\eps_{\spvec{p}'}} \,
\tr (\s{p} + m) \Gamma^\mu(p,q_1,-p') 
\\\times \,
(\s{p}' - m)  \overline{\Gamma}^\nu(p,q_2,-p')
\end{multline}
($p^2=p'^2=m^2$), which is in general not simply contracted with $\eps_\mu \eps^*_\nu$ and $q^\mu_1\neq q^\mu_2$. Nevertheless, it can be related to the polarization operator by considering the following linear combinations
\begin{gather}
\mathfrak{M}_{(\pm)}^{\mu\nu}(q_1,q_2) = \frac{1}{2} \lsb \mathfrak{M}^{\mu\nu}(q_1,q_2) \pm \mathfrak{M}^{\nu\mu}(q_2,q_1) \rsb, 
\end{gather}
where $\mathfrak{M}_{(+)}^{\mu\nu}(q_1,q_2)$ is purely real and $\mathfrak{M}_{(-)}^{\mu\nu}(q_1,q_2)$ purely imaginary, as $[\mathfrak{M}^{\mu\nu}(q_1,q_2)]^* = \mathfrak{M}^{\nu\mu}(q_2,q_1)$  [compare with Eq. (\ref{eqn:paircreation_polarizationoperatortracecc})]. After similar manipulations as in Appendix \ref{sec:cuttingrulesappendix} we obtain 
\begin{gather}
\begin{aligned}
\mathfrak{M}_{(+)}^{\mu\nu}(q_1,q_2)
&=
\Im [\mc{P}^{\mu\nu}(q_1,q_2) + \mc{P}^{\nu\mu}(q_2,q_1)],
\\
\mathfrak{M}_{(-)}^{\mu\nu}(q_1,q_2)
&=
-\I \Re [\mc{P}^{\mu\nu}(q_1,q_2) - \mc{P}^{\nu\mu}(q_2,q_1)]
\end{aligned}
\end{gather}
for $q_i^\lminus > 0$ (which is necessary if the electron and the positron are real, i.e. for $p^\lminus,p'^\lminus > 0$).

\section{Exact photon wave function}
\label{sec:exactphotonwavefunction}

It is shown in Sec. \ref{sec:numericalresults} that the quantity $\probsym(q)$ [see Eq. (\ref{eqn:paircreation_decayratefinalAcopy})] may become larger than unity for next-generation laser parameters. Correspondingly, in this regime the common interpretation of $\probsym(q)$ as pair-production probability must be modified. The reason for this at first sight unphysical result is the fact that we neglected the exponential decay of the photon (and of the electron) wave function during the evaluation of the Feynman diagram shown in Fig. \ref{fig:pair}. As the photon is unstable in the external field, the phase of its exact wave function must contain an imaginary part to ensure a unitary time evolution. In the realm of quantum field theory unitarity leads to the optical theorem [see Eq. (\ref{eqn:paircreation_opticaltheoremmaintext})] and the decay of the wave function is obtained naturally from radiative corrections to the exact external photon line \cite{landau_quantum_1981,schwinger_gauge_1951}. As we will show now, the decay of the exact photon wave function must be taken into account self-consistently for the calculation of the pair-creation probability if $\probsym(q) \gtrsim 1$ (see \cite{dinu_vacuum_2014} for an equivalent approach). 

In vacuum the wave function of an incoming (outgoing) photon with four-momentum $q^\mu$ is given by \cite{peskin_introduction_2008,landau_quantum_1981}
\begin{gather}
\Phi_{q,j}^{\text{in}\mu}(x) = \eps_j^\mu e^{-\I qx},
\quad 
\Phi^{*\text{out}\mu}_{q,j}(x) = \eps_j^{*\mu} e^{\I q x}
\end{gather}
($j=1,2$ labels the polarization state). However, inside a background field radiative corrections affect also the external lines (even after renormalization \cite{weinberg_quantum_1995}) and the exact wave function obeys the Schwinger-Dyson equation \cite{landau_quantum_1981,schwinger_gauge_1951}
\begin{gather}
\label{eqn:schwingerequations}
\begin{gathered}
\begin{aligned}
-\del^2 \Phi_{q}^{\text{in}\mu}(x) =& \int d^4y \, P^{\mu\nu}(x,y) \Phi^{\text{in}}_{q\nu}(y),
\\
-\del^2 \Phi^{*\text{out}\mu}_{q}(x) =& \int d^4y \, \Phi^{*\text{out}}_{q\nu}(y) P^{\nu\mu}(y,x),
\end{aligned}
\end{gathered}
\end{gather}
where $P^{\mu\nu}(x,y)$ denotes the polarization operator in position space. It is related to the polarization operator in momentum space (see Appendix \ref{sec:poloppaperresult}) via
\begin{gather}
\label{eqn:poloppositionvsmomentum}
\begin{aligned}
P^{\mu\nu}(x,y)
&=
\int \frac{d^4q_1 d^4q_2}{(2\pi)^8} \, e^{-\I q_2 x} \, \mc{P}^{\mu\nu}(-q_2,-q_1) \, e^{\I q_1 y},
\\
\mc{P}^{\mu\nu}(q_1,q_2)
&=
\int d^4x d^4y \,  e^{-\I q_1 y} \, P^{\mu\nu}(y,x) \, e^{\I q_2 x} 
\end{aligned}
\end{gather}
[note that we use for $P^{\mu\nu}(x,y) = P^{\nu\mu}(y,x)$ the Schwinger notation (particle propagation from $y$ to $x$) \cite{schwinger_gauge_1951}, while in $\mc{P}^{\mu\nu}(q_1,q_2)=\mc{P}^{\nu\mu}(-q_2,-q_1)$ the incoming momentum is denoted by $q_1$]. As the vacuum part does not change the photon wave function (after renormalization is carried out), we will consider only the field-dependent part of the polarization operator in the following. 

We point out that now even for a plane-wave field the states for incoming and outgoing particles are not equivalent anymore, as the quantum loop enters differently into the equations (the corresponding Schwinger-Dirac equation for an electron is discussed in \cite{meuren_quantum_2011}).

To solve the Schwinger-Dyson equations [see Eq. (\ref{eqn:schwingerequations})] we use the ansatz $\Phi^{\text{in}\mu}_{q,j}(x) = \exp(-\I qx) \mc{E}^{\text{in}\mu}_{q,j}(kx)$ (incoming photon) and $\Phi^{*\text{out}\mu}_{q,j}(x) = \mc{E}^{*\text{out}\mu}_{q,j}(kx) \exp(\I qx)$ (outgoing photon) and the boundary conditions $\mc{E}^{\text{in}\mu}_{q,j}(-\infty) \to \eps_j^\mu$ and $\mc{E}^{*\text{out}\mu}_{q,j}(+\infty) \to  \eps_j^{*\mu}$ (with a suitable choice for the constant polarization four-vectors $\eps_j^\mu$). We note that the quantum numbers are left unchanged, as they denote the asymptotic momenta (i.e. $q^2 = 0$). However, we obtain $\mathfrak{Q}_j^2 \neq 0$, where the effective four-momentum $\mathfrak{Q}_j^\mu$ is defined as the derivative of the total wave-function phase \cite{dinu_vacuum_2014}, i.e. for in states
\begin{gather}
\label{eqn:photonlocalfourmomentum}
\mathfrak{Q}_j^\mu = - \del^\mu S_{q,j}(x),
\quad
\Phi^{\text{in}\mu}_{q,j}(x) \sim \exp[\I S_{q,j}(x)]
\end{gather}
($\mathfrak{Q}_j^\mu = q^\mu$ to leading order).

From Eq. (\ref{eqn:schwingerequations}) we obtain now the following integro-differential equations
\begin{gather}
\label{eqn:schwingerequationsB}
\begin{gathered}
\begin{aligned}
\I 2kq \, \mc{E}'^{\text{in}\mu}_{q,j}(\phi) &= \int d\phi' \, P^{\mu\nu}_{q}(\phi,\phi') \mc{E}^{\text{in}}_{q,j\nu}(\phi'),
\\
-\I 2kq \, \mc{E}'^{*\text{out}\mu}_{q,j}(\phi) &= \int d\phi' \,  \mc{E}^{*\text{out}}_{q,j\nu}(\phi') \tilde{P}^{\nu\mu}_{q}(\phi',\phi),
\end{aligned}
\end{gathered}
\end{gather}
where we defined the quantities 
\begin{gather}
\label{eqn:lightconepolop}
\begin{aligned}
P^{\mu\nu}_{q}(kx,ky)  &= \int dy^\lplus dy^\lperp \, e^{\I q(x-y)} \, P^{\mu\nu}(x,y),
\\
\tilde{P}^{\nu\mu}_{q}(ky,kx) &= \int dy^\lplus dy^\lperp \, e^{\I q(y-x)} \, P^{\nu\mu}(y,x)
\end{aligned}
\end{gather}
(we use the same notation for light-cone coordinates as in \cite{meuren_polarization_2013}). 

From now on we focus on the in states and assume that the background field is strong ($\xi\gg1$), i.e. we use the leading-order quasistatic approximation for the polarization operator \cite{meuren_polarization_2013}. To determine the leading-order corrections in $\alpha \chi^{\nfrac{2}{3}} \ll 1$ \cite{ritus_1985}, we can apply a perturbative approach \cite{schwinger_gauge_1951} (i.e. assume that $\mathfrak{Q}_j^\mu = q^\mu$ to leading order) and obtain the following differential equation
\begin{multline}
\label{eqn:schwingerequationsperturbative}
\I 2kq \, \mc{E}'^{\text{in}\mu}_{q,j}(\phi) =  - \big[ \mathfrak{p}_1(\phi) \, \Lambda_1^\mu \Lambda_1^\nu + \mathfrak{p}_2(\phi)\, \Lambda_2^\mu \Lambda_2^\nu 
\\+ \mathfrak{p}_3(\phi) \frac{q^\mu q^\nu}{m^2} \big] \mc{E}^{\text{in}}_{q,j\nu}(\phi),
\end{multline}
where \cite{meuren_polarization_2013}
\begin{gather}
\begin{aligned}
\mathfrak{p}_1(\phi)
&=
\phantom{-}\alpha\, \frac{m^2}{3\pi} \int_{-1}^{+1} dv \, (w-1) \frac{f'(\tilde{x})}{\tilde{x}},\\
\mathfrak{p}_2(\phi)
&=
\phantom{-}\alpha\, \frac{m^2}{3\pi} \int_{-1}^{+1} dv \, (w+2) \frac{f'(\tilde{x})}{\tilde{x}},\\
\mathfrak{p}_3(\phi) 
&= - \alpha\, \frac{m^2}{\pi} \int_{-1}^{+1} dv \, \frac{f_1(\tilde{x})}{w}
\end{aligned}
\end{gather}
[$\tfrac{1}{w} = \tfrac14(1-v^2)$, $\tilde{x} = [\nfrac{w}{\abs{\chi(\phi)}}]^{\nfrac{2}{3}}$, $\chi(\phi) = \chi \psi'(\phi)$ and the Ritus functions are defined in Eq.~(\ref{eqn:polop_ritusfunctions})]. As the incoming photon is initially transversely polarized (i.e. in the plane spanned by $\Lambda_j^\mu$, $j=1,2$), the term proportional to $\mathfrak{p}_3(\phi)$ can be ignored.

Finally, we obtain for the exact incoming photon wave function the solutions
\begin{gather}
\label{eqn:exactphotonwavefunction}
\Phi^{\text{in}\mu}_{q,j}(x) =  \Lambda_j^\mu \exp\bigg[-\I qx -\I \frac{1}{2kq} \int_{-\infty}^{kx} d\phi' \, \mathfrak{p}_j(\phi')\bigg]
\end{gather}
($j=1,2$, valid for $\xi \gg 1$ and as long as $\alpha \chi^{\nfrac{2}{3}} \ll 1$ \cite{ritus_1985}, the same calculation can also be applied for the outgoing photon wave function). From Eq. (\ref{eqn:exactphotonwavefunction}) we conclude that the effective photon four-momentum is given by [see Eq. (\ref{eqn:photonlocalfourmomentum})] 
\begin{gather}
\mathfrak{Q}_j^\mu(\phi) = q^\mu + k^\mu \frac{1}{2kq} \mathfrak{p}_j(\phi)
\end{gather} 
and the well-known expression for the square of the photon mass $\mathfrak{Q}_j^2$ is recovered \cite{ritus_radiative_1972,batalin_greens_1971,ritus_radiative_1970,narozhnnyi_radiative_1969} (for other discussions of refractive indices and birefringence see e.g. \cite{dinu_vacuum_2014,dittrich_probingquantum_2000,affleck_photon_1988,baier_interaction_1975,bialynicka-birula_nonlinear_1970,baier_1967} and the reviews \cite{di_piazza_extremely_2012,marklund_nonlinear_2006}).

Due to the imaginary part of the polarization operator the photon wave function decays exponentially (as expected for an unstable particle). During the calculation of the pair-creation probability (see Appendix \ref{sec:paircreationappendix}) the exponential decay of the wave function must be considered as soon as $\probsym(q)$ [see Eq. (\ref{eqn:paircreation_decayratefinalAcopy})] becomes of order one.

The total probability that a photon with polarization four-vector $\eps_j^\mu = \Lambda_j^\mu$ does not decay inside the laser pulse is obtained by evaluating the square of the exact wave function at $kx\to\infty$, it is given by (see also \cite{brezin_pair_1970})
\begin{gather}
\label{eqn:photonsurvivalprobability}
W_{q,j}^{\text{s}} = \exp\bigg\{\frac{1}{kq} \int_{-\infty}^{+\infty} d\phi' \, \Im [\mathfrak{p}_j(\phi')] \bigg\}.
\end{gather}
As the imaginary part of the polarization operator is related to the pair-creation diagram (without radiative corrections) via the optical theorem [see Eq. (\ref{eqn:paircreation_opticaltheoremmaintext})]
\begin{gather}
\probsym_j(q)
=
\frac{1}{kq} \, \Im \, [\Lambda_{j\mu} \Lambda_{j\nu} \, \Pi^{\mu\nu}(q,q)],
\end{gather}
the survival probability can be expressed as 
\begin{gather}
\label{eqn:photonsurvivalprobabilityB}
W_{q,j}^{\text{s}} = \exp[-\probsym_j(q)] \approx 1 - \probsym_j(q)
\end{gather}
[the last relation holds only if $\probsym_j(q)$ is much smaller than unity]. Thus, $\probsym_j(q)$ must be interpreted as the decay exponent for the photon wave function if it becomes large [nevertheless, we call $\probsym_j(q)$ the probability for pair creation in the following].

\section{Pair-creation probability}
\label{sec:paircreationprobability}

In order to obtain a compact expression for the total nonlinear Breit-Wheeler pair-creation probability $\probsym(q)$ [see Eq.~(\ref{eqn:paircreation_decayratefinalAcopy})], we apply the optical theorem [see Eq.~(\ref{eqn:paircreation_opticaltheoremmaintext})] to the double-integral representation for the field-dependent part of the polarization operator given in Eq.~(\ref{eqn:polarizationoperator_realincoming_Wfunctions}). For a single on-shell photon with four-momentum $q^\mu$ and polarization four-vector $\eps^\mu$, colliding with a plane-wave laser pulse described by the field tensor [see Eq.~(\ref{eqn:fieldtensor})] 
\begin{gather}
F^{\mu\nu}(\phi) 
= 
f_1^{\mu\nu} \psi_1'(\phi)
+
f_2^{\mu\nu} \psi_2'(\phi)
\end{gather}
the result is given by
\begin{multline}
\label{eqn:paircreation_totalprobabilityfinal}
\probsym(q)
=
-\frac{1}{kq} \, \frac{\alpha}{2\pi} \, \int_0^{\infty} \frac{d\varrho}{\varrho} \,  \int_{-\infty}^{+\infty} dy^\lminus
\Im \, \eps_\mu \eps^*_\nu \big[
P_{12} \Lambda_1^\mu \Lambda_2^\nu
\\+
P_{21} \Lambda_2^\mu \Lambda_1^\nu
+
P_{11} \Lambda_1^\mu \Lambda_1^\nu
+
P_{22} \Lambda_2^\mu \Lambda_2^\nu
\big],
\end{multline}
where 
\begin{gather}
\label{eqn:LambdaandQvectors}
\begin{aligned}
\Lambda_1^\mu &= \frac{f_1^{\mu\nu} q_\nu}{kq \sqrt{-a_1^2}},
&
\Lambda_2^\mu &= \frac{f_2^{\mu\nu} q_\nu}{kq \sqrt{-a_2^2}},
\\
\Q_1^\mu &= \frac{k^\mu q_1^2 - q_1^\mu kq}{kq},
&
\Q_2^\mu &= \frac{k^\mu q_2^2 - q_2^\mu kq}{kq},
\end{aligned}
\end{gather}
and the coefficients [see Eq.~(\ref{eqn:polarizationoperator_realincoming_coefficients})] are evaluated at $q_1=q_2=q$, $q^2=0$ [$P_{Q}$ does not contribute for $q\eps=q^2=0$, which can be seen from the definition of the four-vectors $\Q_i^\mu$]. 

The dependence of the pair-creation probability on the external field is determined by the classical intensity parameters $\xi_i$ and the quantum-nonlinearity parameters $\chi_{i}$
\begin{gather}
\label{eqn:paircreation_xichi}
\xi_i 
= 
\frac{\abs{e}}{m} \sqrt{-a_i^2},
\quad
\chi_{i}
=
\frac{\abs{e}\sqrt{qf_i^2q}}{m^3}
=
\eta \, \xi_i,
\end{gather}
where $\eta =  \nfrac{\sqrt{(kq)^2}}{m^2}$. Note that in the general case of elliptic polarization $\xi$ and $\chi$ given in the Introduction represent the characteristic scale of $\xi_i$ and $\chi_i$, respectively (see \cite{meuren_polarization_2013} for details). 

\subsection{Linear polarization}

We consider now the special case of a linearly polarized background field [$\xi = \xi_1$, $\xi_2=0$; $\psi_1(\phi) = \psi(\phi)$, $\psi_2(\phi)=0$; $F^{\mu\nu} = \psi'(\phi) f^{\mu\nu}$, $f^{\mu\nu} = k^\mu a^\nu - k^\nu a^\mu$, $P_{12}=P_{21}=0$]. It is then useful to introduce the following two polarization four-vectors [see Eq.~(\ref{eqn:LambdaandQvectors})]
\begin{gather}
\label{eqn:paircreation_polarizationvectors}
\eps_\parallel^\mu = \Lambda_1^\mu,
\quad
\eps_\perp^\mu = \Lambda_2^\mu.
\end{gather}
They are real, obey $\eps_\parallel^2 = \eps_\perp^2 = -1$, $\eps_\parallel \eps_\perp = 0$ and represent the direction of the electric and the magnetic field of the laser, respectively (in the frame where the incoming photon and the laser pulse collide head-on).

Accordingly, we obtain for the total pair-creation probability in a linearly polarized laser pulse by an on-shell photon with polarization four-vector $\eps_\parallel^\mu$ and $\eps_\perp^\mu$ [see Eq.~(\ref{eqn:paircreation_totalprobabilityfinal})]
\begin{gather}
\label{eqn:paircreation_linpolprobability}
\begin{aligned}
\probsym_\parallel(q) &= 
- \alpha \, \frac{m^2}{kq} \, \frac{1}{2\pi} \, \int_{-\infty}^{+\infty} dy^\lminus \,  \int_0^{\infty} \frac{d\varrho}{\varrho} \,  \Im \, \widetilde{P}_{11},\\
\probsym_\perp(q) &=
- \alpha \, \frac{m^2}{kq} \, \frac{1}{2\pi} \, \int_{-\infty}^{+\infty} dy^\lminus \,  \int_0^{\infty} \frac{d\varrho}{\varrho} \,  \Im \, \widetilde{P}_{22},
\end{aligned}
\end{gather}
where [see Eq.~(\ref{eqn:polarizationoperator_realincoming_coefficients})]
\begin{align}
\widetilde{P}_{11} &= - \frac{i}{\varrho} \frac{kq}{m^2} \Big[ \mc{W}_2(x_1) e^{-\I 4x_1} - \mc{W}_2(x_0)  e^{-\I 4x_0}  \Big] \\\nonumber
&\phantom{= }+  \xi^2 \Big[ \frac{1}{2} V  \mc{W}_0(x_1) + 2 X \mc{W}_1(x_1) \Big] e^{-\I 4x_1},
\nonumber\displaybreak[0]\\\nonumber
\widetilde{P}_{22} &= - \frac{i}{\varrho} \frac{kq}{m^2} \Big[ \mc{W}_2(x_1) e^{-\I 4x_1} - \mc{W}_2(x_0)  e^{-\I 4x_0}  \Big] \\\nonumber
&\phantom{= }+  \xi^2 \frac{1}{2} V  \mc{W}_0(x_1)  e^{-\I 4x_1}
\end{align}
[$V=V_1$, $X=X_{11}$, see Eq.~(\ref{eqn:polop_XVofky}); $x_0$ and $x_1$ are defined in Eq.~(\ref{eqn:x0x1def}) and $\mc{W}_l(x)$ in Eq.~(\ref{eqn:polop_dwfouriertransform})].

\subsection{Strong fields}
\label{sec:paircreation_strongfields}

As the integrals in Eq.~(\ref{eqn:paircreation_linpolprobability}) are oscillatory, it is useful to derive non-oscillatory representations for important limits. In this section we consider a strong ($\xi \gg 1$), linearly polarized background field. In this case the field-dependent contribution to the polarization operator can be written as \cite{meuren_polarization_2013}
\begin{multline}
\label{eqn:polarizationoperatorlinearpolqc}
\I\mc{P}^{\mu\nu}(q_1,q_2)
-
\I\mc{P}^{\mu\nu}_{\Ftilde=0}(q_1,q_2)
= 
\I (2\pi)^3\delta^{(\lminus,\lperp)}(q_1-q_2) 
 \\ \times\,
\int_{-\infty}^{+\infty} dz^\lminus \, e^{i(q_2^\lplus - q_1^\lplus)z^\lminus} 
\bigg[\pi'_1\, \frac{(fq)^\mu (fq)^\nu}{(fq)^2}
\\+ \pi'_2\, \frac{(f^*q)^\mu (f^*q)^\nu}{(f^*q)^2} - \frac{\pi'_3}{q_1 q_2}\, G^{\mu\nu}  \bigg],
\end{multline}
where $f^{*\mu\nu} = \tfrac{1}{2}\eps^{\mu\nu\rho\sigma}f_{\rho\sigma}$ and
\begin{gather}
\label{eqn:polarizationoperatorlinearpolqc_coefficients}
\begin{aligned}
\pi'_1
&=
\phantom{-}\alpha\, \frac{m^2}{3\pi} \int_{-1}^{+1} dv \, (w-1) \bigg[\frac{\abs{\chi(kz)}}{w}\bigg]^{\nicefrac23} f'(\rho),\\
\pi'_2
&=
\phantom{-}\alpha\, \frac{m^2}{3\pi} \int_{-1}^{+1} dv \, (w+2) \bigg[\frac{\abs{\chi(kz)}}{w}\bigg]^{\nicefrac23} f'(\rho),\\
\pi'_3 
&= - \alpha\, \frac{q_1 q_2}{\pi} \int_{-1}^{+1} dv \, \frac{f_1(\rho)}{w}
\end{aligned}
\end{gather}
with $\tfrac{1}{w} = \tfrac14(1-v^2)$, $\rho = \big[\tfrac{w}{\abs{\chi(kz)}}\big]^{\nicefrac23}(1- \tfrac{q_1 q_2}{m^2} \tfrac{1}{w})$, $\chi(kz) = \chi \psi'(kz)$ and $G^{\mu\nu}  = q_2^\mu q_1^\nu - q_1q_2 \, g^{\mu\nu}$. Furthermore, the Ritus functions are defined by \cite{meuren_polarization_2013,ritus_radiative_1972,ritus_1985}
\begin{subequations}
\label{eqn:polop_ritusfunctions}
\begin{multline}
f(x)
= 
\I \int_0^\infty dt \exp\big[-\I\big( t x + \nfrac{t^3}{3}\big)\big] 
\\=
\pi \Gi(x) + \I \pi \Ai(x),
\end{multline}
\begin{gather}
f_1(x)
=
\int_0^\infty \frac{dt}{t} \exp\lb-\I t x \rb \lsb \exp\big(- \I \nfrac{t^3}{3} \big) -1\rsb,
\end{gather}
\end{subequations}
where $\Ai$ and $\Gi$ are the Airy and Scorer function, respectively \cite{olver_nist_2010}. Note that in Ritus' work the normalization of the Airy function is different and also changes [see \cite{ritus_1985}, Appendix C and \cite{nikishov_quantum_1964}, Eq. (B5)]. In Eq.~(\ref{eqn:polarizationoperatorlinearpolqc_coefficients}) the integration variable can be changed using
\begin{gather}
\label{eqn:paircreation_vtowint}
\int_{-1}^{+1} dv = 2 \int_{0}^{1} dv = \int_{4}^{\infty} dw \, \frac{4}{w\sqrt{w(w-4)}}
\end{gather}
(valid for integrands which are even functions of $v$).

The expression given in Eq.~(\ref{eqn:polarizationoperatorlinearpolqc}) was obtained by applying suitable approximations to the triple-integral representation given in Eq. (92) of \cite{meuren_polarization_2013}. It is tempting to apply the same approximations now to the double-integral representation in Eq.~(\ref{eqn:polarizationoperator_realincoming_Wfunctions}). However, the functions $\mc{W}_i$ change over the formation region ($\mc{W}_0$ even has a logarithmic singularity at the origin), which means that this is not possible. 

To determine the pair-creation probabilities we apply the optical theorem given in Eq.~(\ref{eqn:paircreation_opticaltheoremmaintext}) to Eq.~(\ref{eqn:polarizationoperatorlinearpolqc}) and note the identities [see Eq.~(\ref{eqn:LambdaandQvectors})]
\begin{gather}
\begin{aligned}
\Lambda_1^\mu \Lambda_1^\nu
&=
- \frac{(fq)^\mu (fq)^\nu}{(fq)^2},
\\
\Lambda_2^\mu \Lambda_2^\nu
&=
- \frac{(f^*q)^\mu (f^*q)^\nu}{(f^*q)^2}.
\end{aligned}
\end{gather}
Finally, we obtain for the total probability that a single on-shell photon with four-momentum $q^\mu$ and polarization four-vector $\eps_\parallel^\mu$ or $\eps_\perp^\mu$ creates an electron-positron pair inside a strong ($\xi \gg 1$, $\chi < \xi$), linearly polarized laser pulse with field tensor $F^{\mu\nu}(kx) = f^{\mu\nu} \psi'(\phi)$ the following expressions  
\begin{align}
\label{eqn:paircreat_probabilityxilarge}
\probsym_\parallel(q) &= -\alpha \, \frac{m^2}{kq}  \int_{-\infty}^{+\infty} d\phi \int_{-1}^{+1} dv \, \frac{(w-1)}{3} \frac{\Ai'(\tilde{x})}{\tilde{x}},
\nonumber\displaybreak[0]\\
\probsym_\perp(q) &= -\alpha \, \frac{m^2}{kq} \int_{-\infty}^{+\infty} d\phi \int_{-1}^{+1} dv \, \frac{(w+2)}{3} \frac{\Ai'(\tilde{x})}{\tilde{x}},
\end{align}
where $\tfrac{1}{w} = \tfrac14(1-v^2)$, $\tilde{x} = \big[\nfrac{w}{\abs{\chi(\phi)}}\big]^{\nicefrac23}$ and $\chi(\phi) = \chi \psi'(\phi)$ (due to $q\eps=q^2=0$ the coefficient $\pi'_3$ does not contribute). We point out that Eq.~(\ref{eqn:paircreat_probabilityxilarge}) holds for an arbitrary shape of the plane-wave background field ($\chi$ should be such that $\alpha\chi^{\nfrac{2}{3}} \ll 1$, otherwise perturbation theory with respect to the radiation field is expected to break down \cite{ritus_1985,di_piazza_extremely_2012}). As the formation region is small for $\xi \gg 1$, the total pair-creation probability given in Eq.~(\ref{eqn:paircreat_probabilityxilarge}) consists essentially of the probability to create a pair inside a constant-crossed field [see \cite{ritus_radiative_1972}, Eq. (64) and \cite{ritus_1985}, Chap. 5, Eq. (60); see also \cite{baier_electromagnetic_1994}], integrated over the pulse shape [$\chi(\phi)$ represents the instantaneous value of the quantum-nonlinearity parameter] \cite{ritus_1985}. 

For comparison with the literature we consider now the monochromatic limit of Eq.~(\ref{eqn:paircreat_probabilityxilarge}), i.e. $\psi'(\phi) = \sin(\phi)$ and a counterpropagating photon. As the wave is periodic, we can split the integral in $\phi$ and consider only a single half-cycle (i.e. $\phi \in [0,\pi]$). As the photon is counterpropagating, it passes this half-cycle in the time $\nfrac{T}{4}$, where the laser period is given by $T = \nfrac{2\pi}{\omega}$. Correspondingly, the rate for pair creation by a single photon inside a strong ($\xi\gg 1$), linearly polarized, monochromatic plane wave is given by
\begin{align}
\label{eqn:paircreat_probabilityxilarge_monochromatic}
\probsym_\parallel(q) &= -\alpha \, \frac{m^2}{q^0}  \frac{1}{\pi} \int_{0}^{\pi} d\phi \int_{-1}^{+1} dv \, \frac{(w-1)}{3} \frac{\Ai'(x_m)}{x_m},
\nonumber\displaybreak[0]\\
\probsym_\perp(q) &= -\alpha \, \frac{m^2}{q^0} \frac{1}{\pi} \int_{0}^{\pi} d\phi \int_{-1}^{+1} dv \, \frac{(w+2)}{3} \frac{\Ai'(x_m)}{x_m},
\end{align}
where now $x_m = \big[\nfrac{w}{\abs{\chi_m(\phi)}}\big]^{\nicefrac23}$, $\chi_m(\phi) = \chi \sin(\phi)$ [$\tfrac{1}{w} = \tfrac14(1-v^2)$]. Equation (\ref{eqn:paircreat_probabilityxilarge_monochromatic}) coincides with the result obtained in \cite{ritus_1985} [Chap. 3, Eq. (35) and Chap. 5, Eq. (60)]. It is also in agreement with the results obtained in \cite{baier_interaction_1975}.

\subsection{Small quantum parameter}

For $\chi\ll 1$ the pair-creation probability is exponentially suppressed. This becomes obvious from the asymptotic expansion of the Airy function \cite{olver_nist_2010}
\begin{gather}
\Ai'(x) 
\sim 
- \frac{x^{\nicefrac14} \, e^{-\frac23 x^{3/2}}}{2\sqrt{\pi}}.
\end{gather}
In this regime we can approximately evaluate the integrals in Eq.~(\ref{eqn:paircreat_probabilityxilarge}), resulting in a compact expression for the pair-creation probability. As the pair-creation probability is exponential suppressed, only the region around the peak of the field strength contributes to the integral in $\phi$. Furthermore, using Eq.~(\ref{eqn:paircreation_vtowint}), we see that the integral in $w$ is formed around $w=4$. Correspondingly, we can use 
\begin{gather}
\int_{4}^{\infty} dw \, \frac{1}{\sqrt{w-4}} e^{-xw} = e^{-4x} \sqrt{\frac{\pi}{x}}
\end{gather}
(assuming $x>0$) to evaluate the integral in $w$ approximately. Assuming that $\abs{\psi'(\phi)} \approx \abs{\sin(\phi)}$ close to a field peak, the contribution from this peak can be approximately taken into account using
\begin{gather}
\int_{0}^{\pi} d\phi \, e^{-\nfrac{x}{\sin(\phi)}}  
\approx 
\int_{-\infty}^{+\infty} d\phi \, e^{-x (1+\nfrac{\phi^2}{2})} 
=
e^{-x} \sqrt{\frac{2\pi}{x}}
\end{gather}
(assuming $x>0$; for different peak shapes this relation must be modified accordingly). Combining everything, the pair-creation probability within a single peak of a linearly polarized, plane-wave laser field is in the regime $\xi \gg 1$, $\chi \ll 1$ given by
\begin{gather}
\label{eqn:paircreat_probabilityxilargechimoderate}
\probsym_\parallel(q) = \alpha \, \frac{m^2}{kq}  \frac{3\sqrt{\pi}}{8} \lb \frac{\chi}{2} \rb^{\nfrac{3}{2}} e^{-\nfrac{8}{(3\chi)}},
\end{gather}
$\probsym_\perp(q) = 2\probsym_\parallel(q)$. From Eq. (\ref{eqn:paircreat_probabilityxilargechimoderate}) the pair-creation rate inside monochromatic fields can be obtained similar as above [see Eq. (\ref{eqn:paircreat_probabilityxilarge_monochromatic})]. To this end we consider again a photon counterpropagating with a monochromatic wave. A counterpropagating photon passes four field maxima during the time of one laser period $T = \nfrac{2\pi}{\omega}$. Correspondingly, the pair-creation rate for a single photon is given by (linear polarization, $\xi \gg 1$, $\chi \ll 1$)
\begin{gather}
\label{eqn:paircreat_probabilityxilargechimoderate_monochromaticwave}
\probsym_\parallel(q) = \alpha \, \frac{m^2}{q^0}  \frac{3}{8\sqrt{\pi}} \lb \frac{\chi}{2} \rb^{\nfrac{3}{2}} e^{-\nfrac{8}{(3\chi)}},
\end{gather}
$\probsym_\perp(q) = 2\probsym_\parallel(q)$. The result agrees with \cite{ritus_1985}, Chap. 3, Eq. (33) (see also \cite{reiss_absorption_1962}).

\section{Numerical results}
\label{sec:numericalresults}

\begin{figure}
\centering
\includegraphics{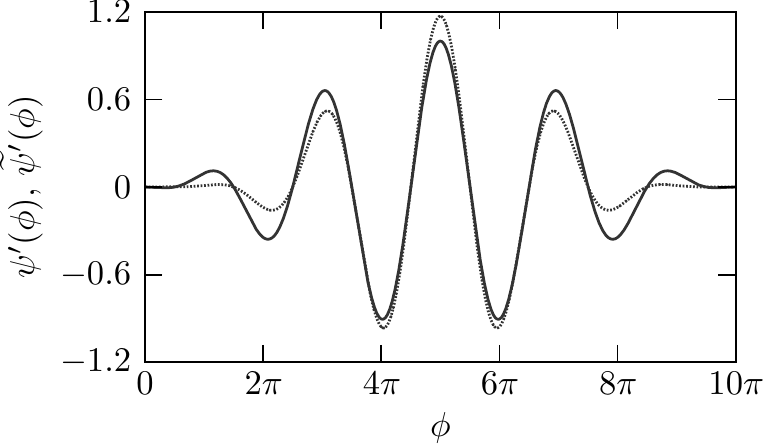}
\caption{\label{fig:pulseshapecompare}In the numerical calculations we considered either a laser pulse with $\sin^2$ (solid line) or with $\sin^4$ envelope (dotted line), plotted here for $N=5$ cycles and a CEP of $\phi_0=-\nfrac{\pi}{2}$ [see Eq.~(\ref{eqn:pulseshape})]. As the $\sin^4$ pulse falls off faster at the edges, it must have a higher peak strength in comparison with a $\sin^2$ envelope in order to describe a pulse with the same total energy.}
\end{figure}

\begin{figure}
\centering
\includegraphics{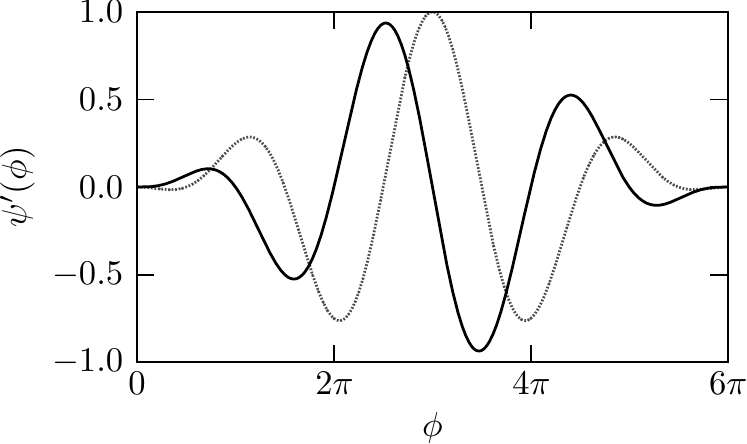}
\caption{\label{fig:pulseshape}Laser pulse with $\sin^2$-envelope [see Eq.~(\ref{eqn:pulseshape})] for different carrier-envelope phases [$\phi_0=0$ (solid line) and $\phi_0=-\nfrac{\pi}{2}$ (dotted line), $N=3$ cycles]. Depending on the CEP we obtain either one strong peak or two peaks with slightly less strength. In the regime where pair creation is exponentially suppressed the first situation is favorable.}
\end{figure}

\highlight{For the numerical calculations we considered a linearly polarized laser pulse with one of the following pulse-shapes [see Fig.~\ref{fig:pulseshapecompare} and Eq.~(\ref{eqn:fieldtensor})]
\begin{gather}
\label{eqn:pulseshape}
\begin{aligned}
\psi'(\phi) 			&= \phantom{\mc{R}} \sin^2[\nfrac{\phi}{(2N)}] \, \sin(\phi + \phi_0),
\\
\widetilde{\psi}'(\phi) &=  \mc{R} \sin^4[\nfrac{\phi}{(2N)}] \, \sin(\phi + \phi_0)
\end{aligned}
\end{gather}
for $\phi \in [0,2\pi N]$ and zero otherwise [if not specified explicitly, we use $\psi'(\phi)$]. Here $N$ characterizes the number of cycles in the pulse and $\phi_0$ the carrier-envelope phase (CEP, see Fig. \ref{fig:pulseshape}). The scaling parameter $\mc{R}$ is chosen such that
\begin{gather}
\int_0^{2\pi N} d\phi \, [\psi'(\phi)]^2 = \int_0^{2\pi N} d\phi \, [\widetilde{\psi}'(\phi)]^2,
\end{gather}
as in most experiments the total energy in the pulse is fixed and only the pulse shape itself may change (see Fig. \ref{fig:pulseshapecompare}). As long as $N$ is an integer and $N\geq 2$ for $\psi'$ and $N\geq 3$ for $\widetilde{\psi}'$, the above shape functions describe a laser pulse without dc component, i.e.
\begin{gather}
\int_0^{2\pi N} d\phi \, \psi'(\phi) = \int_0^{2\pi N} d\phi \, \widetilde{\psi}'(\phi) = 0
\end{gather}
and the energy in the pulse is independent of the CEP.}

Beside the pulse shape parametrized by $N$ and $\phi_0$ we have to chose the classical intensity parameter $\xi$ and the quantum-nonlinearity parameter $\chi = \eta \xi$ [$\eta = \nfrac{\sqrt{(kq)^2}}{m^2}$, see Eq.~(\ref{eqn:paircreation_xichi})]. For a laser pulse with central angular frequency $\omega$ and peak field amplitude $E_0$ we obtain $\xi=\nfrac{|e|E_0}{(m\omega)}$. The quantum-nonlinearity parameter is given by $\chi=2(\nfrac{\omega_\gamma}{m}) (\nfrac{E_0}{E_{cr}})$ if the photon and the laser pulse collide head-on ($\omega_\gamma$ denotes the energy of the incoming photon). 

As the pair-creation probability is exponentially suppressed for $\chi \ll 1$ [see Eq.~(\ref{eqn:paircreat_probabilityxilargechimoderate})], we are mainly interested in the nonlinear quantum regime where $\chi \gtrsim 1$. Existing optical petawatt laser systems reach already $\xi\sim 100$ \cite{yanovsky_ultra_2008} and photon energies $\sim \unit[1]{GeV}$ are obtainable via Compton backscattering either at conventional facilities like SPring-8 \cite{muramatsu_development_2014} or by using laser wakefield accelerators \cite{leemans_gev_2006,esarey_physics_2009,phuoc_all-optical_2012,wang_quasi-monoenergetic_2013,powers_quasi-monoenergetic_2014}. Hence, it is possible to reach the regime $\chi \gtrsim 1$ with presently available technology.

\begin{figure}
\centering
\includegraphics{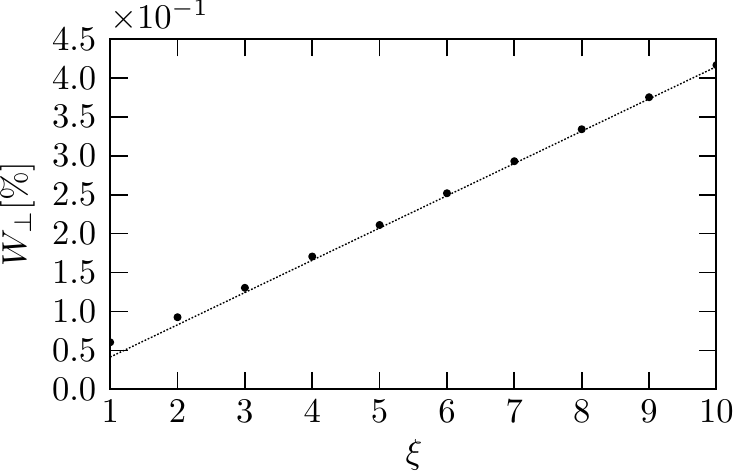}
\caption{\label{fig:plot_comparison}Total pair-creation probability for a single photon inside a linearly polarized laser pulse ($\phi_0=0$, $N=5$, $\chi=1$). The full numerical calculation [black dots, see Eq.~(\ref{eqn:paircreation_linpolprobability})] is compared with the constant-crossed field approximation [valid for $\xi \gg 1$, dotted line, see Eq.~(\ref{eqn:paircreat_probabilityxilarge})].}
\end{figure}

\begin{figure}
\centering
\includegraphics{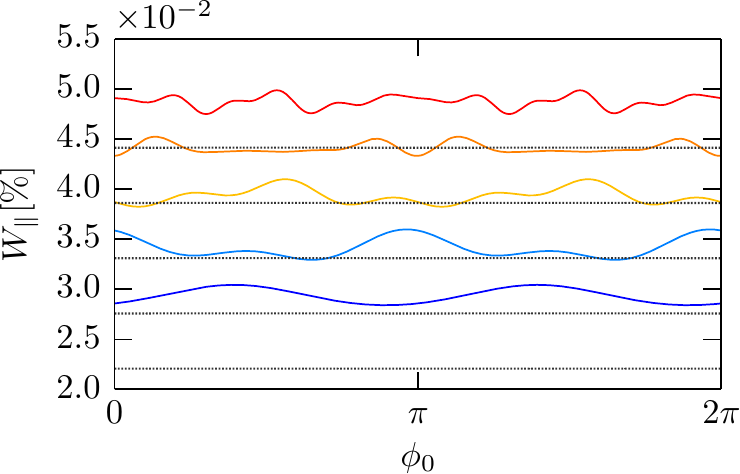}
\caption{\label{fig:plot_smallxicep}Comparison between the full numerical calculation of the pair-creation probability [solid lines, see Eq.~(\ref{eqn:paircreation_linpolprobability})] and the constant-crossed field approximation [valid for $\xi \gg 1$, dotted lines, see Eq.~(\ref{eqn:paircreat_probabilityxilarge})] for different values of $\xi$ ($1.0$, $1.25$, $1.5$, $1.75$ and $2.0$, lower to upper curve, $N=5$, $\chi=1$). Besides underestimating the probability, the strong-field approximation fails to reproduce the correct CEP dependence. This is expected, as the formation region is large for $\xi \sim 1$ and nonlocal properties of the pulse play an important role.}
\end{figure}

In the following we will not consider the influence of the incoming photon wave packet and set $\probsym = \probsym(q)$, see Eq. (\ref{eqn:paircreation_decayratefinalA}). For $\xi\gg 1$ the total pair-creation probability can be calculated using Eq.~(\ref{eqn:paircreat_probabilityxilarge}) without further numerical difficulties, as the integrals are non-oscillatory. To verify the validity of Eq.~(\ref{eqn:paircreat_probabilityxilarge}), we have compared it with the general expression given in Eq.~(\ref{eqn:paircreation_linpolprobability}) (the oscillatory integrals have been evaluated numerically as explained in Appendix~\ref{sec:numericalintegration}). The result is shown in Fig.~\ref{fig:plot_comparison}. Already for $\xi\lesssim 10$ both equations are in good agreement. \highlight{However, for $\xi \sim 1$ the constant-crossed field approximation fails to predict the CEP dependence of the probability (see Fig. \ref{fig:plot_smallxicep}). As the formation region scales as $\nfrac{1}{\xi}$, the global structure of the pulse within the formation region itself [which is not included in the constant-crossed field approximation, see Eq. (\ref{eqn:paircreat_probabilityxilarge})] becomes important at $\xi \sim 1$.} 

From now on we consider the experimentally interesting regime $\xi \gg 1$ and use Eq.~(\ref{eqn:paircreat_probabilityxilarge}) to determine the total pair-creation probability. Furthermore, we compare the two different pulse shapes given in Eq. (\ref{eqn:pulseshape}) (solid lines are calculated using the $\sin^2$-envelope, dotted lines using the $\sin^4$-envelope). \highlight{In general, the results do not depend strongly on the pulse shape. However, in the regime where pair creation is exponentially suppressed ($\chi \ll 1$), the $\sin^4$-envelope is favorable, as it implies a higher peak field strength (see Fig. \ref{fig:pulseshapecompare}).}

\begin{figure}
\centering
\includegraphics{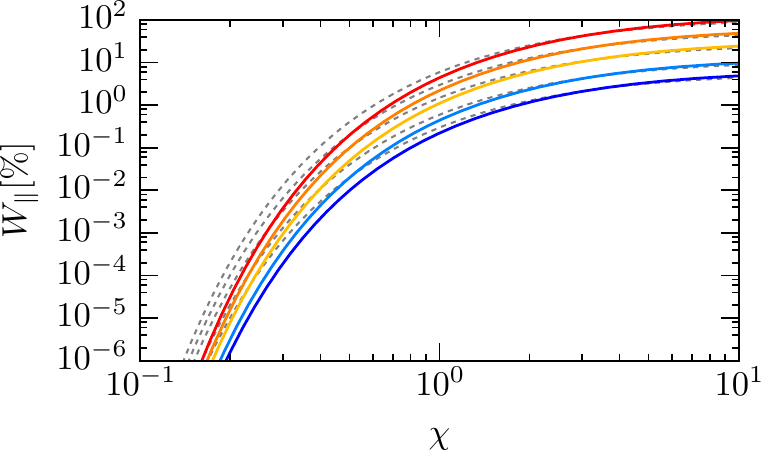}
\caption{\label{fig:plot_xichi}Dependence of the pair-creation probability on $\xi$ and $\chi$ ($\phi_0=0$, $N=5$). For $\xi \gtrsim 1$ the pair-creation probability increases linearly with $\xi$ (we plotted the values $\xi=10$, $20$, $50$, $100$ and $200$, lower to upper curve). In the regime $\chi \ll 1$ pair creation is exponentially suppressed [pulse shape $\psi'(\phi)$ (solid lines) and $\widetilde{\psi}'(\phi)$ (dotted lines), see Eq. (\ref{eqn:pulseshape})].}
\end{figure}

\begin{figure}
\centering
\includegraphics{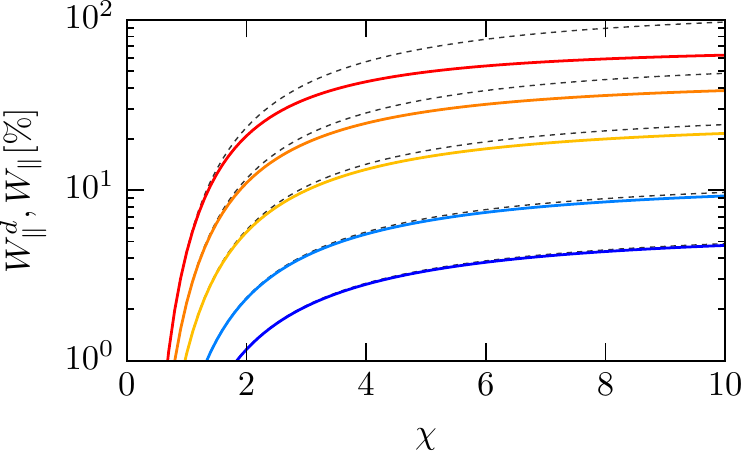}
\caption{\label{fig:plot_xichicomparison}Comparison between the photon-decay probability $W^{d}_{\parallel}$ (solid lines) and the pair-creation probability $\probsym_{\parallel}$ obtained from the leading-order Feynman diagram shown in Fig.~\ref{fig:pair}, i.e. without including radiative corrections (dashed lines). As long as the probability is small, both quantities agree. However, for $\probsym_{\parallel} \sim 1$ it is important to note that $\probsym_{\parallel}$ represents the decay exponent of $W^{d}_{\parallel}$ [see Eq. (\ref{eqn:photonsurvivalprobabilityB}), the parameters are as in Fig. \ref{fig:plot_xichi}, the pulse shape is given by $\psi'(\phi)$].}
\end{figure}

In Fig.~\ref{fig:plot_xichi} we plot the total pair-creation probability as a function of the parameters $\xi$ and $\chi$. For $\xi \gtrsim 1$ it scales linear in $\xi$ due to the phase-space prefactor $\nfrac{m^2}{kq}$ [see Eq.~(\ref{eqn:paircreat_probabilityxilarge})] and only the dependence on $\chi$ is nontrivial \cite{ritus_1985}. Around $\chi \sim 1$ we leave the region of exponentially suppression [see Eq.~(\ref{eqn:paircreat_probabilityxilargechimoderate})] and the pair-creation probability becomes sizable \cite{ritus_1985}. 

\highlight{As explained in Sec. \ref{sec:exactphotonwavefunction}, the quantities $\probsym_{\parallel,\perp}$ only represent the total pair-creation probability as long as they are much smaller than unity. In general, one has to consider the total probability for the decay of a photon with a given polarization [see Eq. (\ref{eqn:photonsurvivalprobabilityB})]
\begin{gather}
W^{d}_{\parallel,\perp} = 1 - W_{\parallel,\perp}^{\text{s}} = 1 - \exp[-\probsym_{\parallel,\perp}] \approx \probsym_{\parallel,\perp}.
\end{gather}
In Fig. \ref{fig:plot_xichicomparison} we have compared both quantities to show when this difference becomes relevant. We point out that the decay of the photon is necessarily accompanied by the creation of at least one electron-positron pair.}

\begin{figure}
\centering
\includegraphics{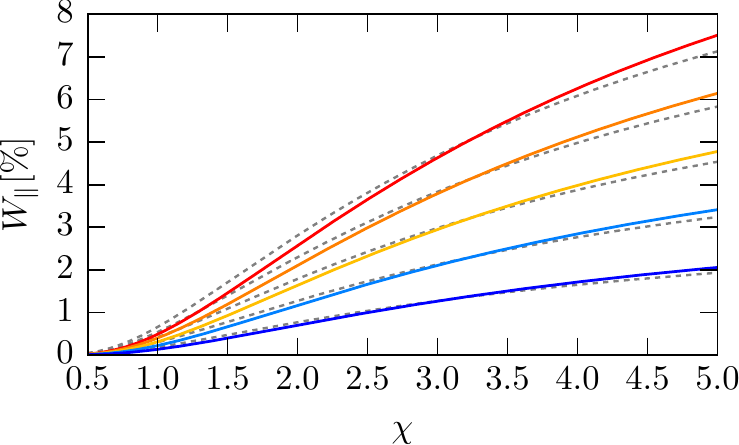}
\caption{\label{fig:plot_Nplot}The scaling of the pair-creation probability is roughly linear in the pulse length (we plotted $N=3$, $5$, $7$, $9$ and $11$, lower to upper curve; $\phi_0=0$, $\xi=10$) [pulse shape $\psi'(\phi)$ (solid lines) and $\widetilde{\psi}'(\phi)$ (dotted lines), see Eq. (\ref{eqn:pulseshape})].}
\end{figure}

The dependence of the total pair-creation probability on the pulse length $N$ is shown in Fig.~\ref{fig:plot_Nplot}. As expected, the scaling in the pulse length is roughly linear in the regime $\chi \sim 1$.

\begin{figure}
\centering
\includegraphics{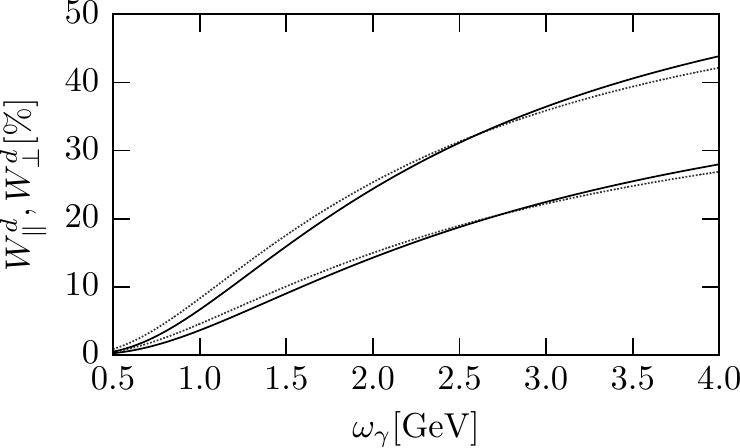}
\caption{\label{fig:plot_photonenergy}Total photon-decay probability during the head-on collision between a gamma photon with energy $\omega_\gamma$ and a linearly polarized laser pulse with the following parameters: $\phi_0=0$, $N=5$, $\omega=\unit[1.55]{eV}$, $\xi=100$ [pulse shape $\psi'(\phi)$ (solid lines) and $\widetilde{\psi}'(\phi)$ (dotted lines), see Eq. (\ref{eqn:pulseshape})]. For parallel polarization ($W^d_\parallel$) the probabilities are smaller than for orthogonal polarization ($W^d_\perp$).}
\end{figure}

In Fig.~\ref{fig:plot_photonenergy} we have plotted the parameter regime accessible by combining a petawatt laser system ($\xi=100$) with a $\unit{GeV}$ photon source. Accordingly, it is possible to obtain a large pair-creation yield even with a limited number of highly energetic photons. As expected from Eq. (\ref{eqn:paircreat_probabilityxilargechimoderate}), the pair-creation probability for perpendicular polarization ($W_\perp$) is roughly twice as large as for parallel polarization ($W_\parallel$). Correspondingly, $W_\parallel$ can  be considered as a lower bound (which is the reason why we mainly focused on this polarization).

\begin{figure}
\centering
\includegraphics{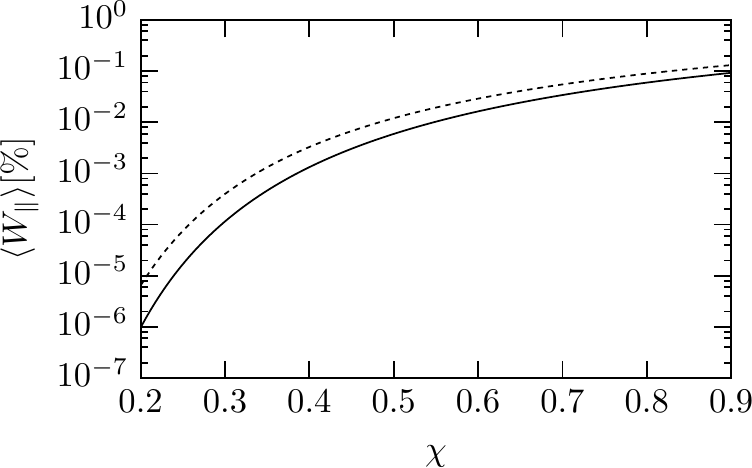}
\caption{\label{fig:plot_cepaverage}The total pair-creation probability as a function of the quantum-nonlinearity parameter $\chi$ averaged over the CEP phase [see Eq.~(\ref{eqn:paircreation_relativecep}), $\xi=10$ and $N=3$]. The solid and the dotted line correspond to the pulse shape $\psi'(\phi)$ and $\widetilde{\psi}'(\phi)$, respectively [see Eq. (\ref{eqn:pulseshape})].}
\end{figure}

\begin{figure*}[bth]
\centering
\subfloat{\includegraphics{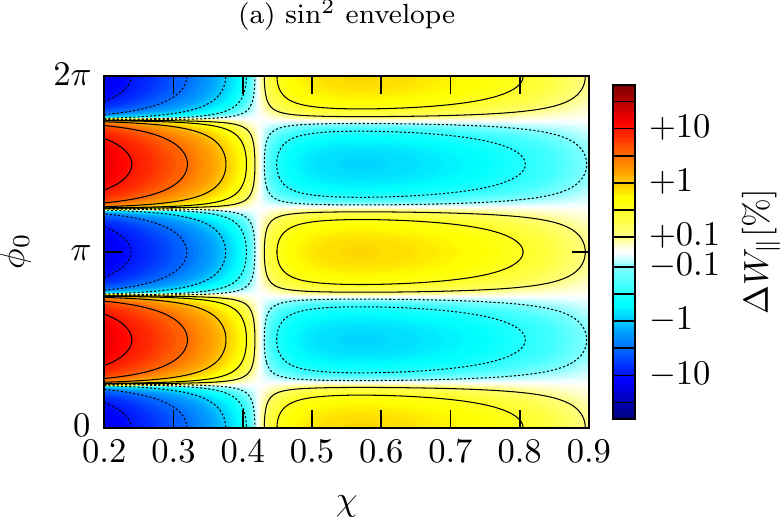}}
\hspace*{10pt}
\subfloat{\includegraphics{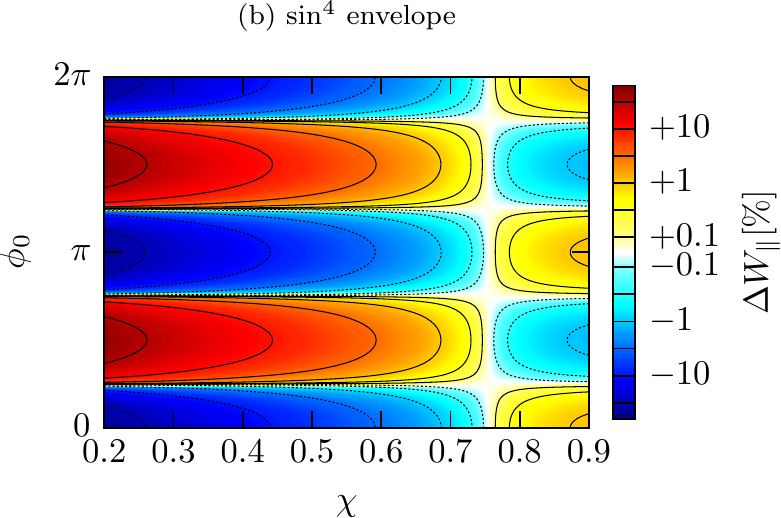}}
\caption{\label{fig:plot_cep}The relative pair-creation probability [see Eq.~(\ref{eqn:paircreation_relativecep})] as a function of the quantum-nonlinearity parameter $\chi$ and the CEP $\phi_0$ ($\xi=10$ and $N=3$, for the CEP-averaged probability see Fig. \ref{fig:plot_cepaverage}). We used the pulse shape $\psi'(\phi)$ for the left and $\widetilde{\psi}'(\phi)$ for the right plot [see Eq. (\ref{eqn:pulseshape})]. The dependence on the CEP is quite pronounced for $\chi \ll 1$, where, however, the total probability is strongly suppressed (the intermediate color levels are located at $10^{-\nfrac{1}{2}} \approx 0.32$, $10^{\nfrac{1}{2}} \approx 3.2$ and $10^{\nfrac{3}{2}} \approx 32$).}
\end{figure*}

Due to the fact that the pair-creation probability is exponentially suppressed for $\chi \ll 1$, it depends very sensitively on the maximum field strength in this regime and large CEP effects can be expected. To investigate them, we introduce the CEP-averaged pair-creation probability
\begin{subequations}
\label{eqn:paircreation_relativecep}
\begin{gather}
\langle \probsym_{\parallel,\perp} \rangle = \frac{1}{2\pi} \int_{0}^{2\pi} d\phi_0 \, \probsym_{\parallel,\perp}(\phi_0)
\end{gather}
and the relative deviation
\begin{gather}
\Delta \probsym_{\parallel,\perp}(\phi_0)  = \frac{\probsym_{\parallel,\perp}(\phi_0)-\langle \probsym_{\parallel,\perp} \rangle}{\langle \probsym_{\parallel,\perp} \rangle}.
\end{gather}
\end{subequations}
They are plotted in Fig. \ref{fig:plot_cepaverage} and Fig. \ref{fig:plot_cep}, respectively, for a short pulse ($N=3$) of moderate intensity ($\xi=10$). For $\chi \approx 0.2$ the relative CEP effect is of the order of $10\%$, but many photons are needed to produce a sufficient amount of electron-positron pairs. In the regime where pair creation is likely ($\chi \sim 1$), the CEP effect for the total pair-creation probability is very small (we point out that this prediction could be changed by higher-order corrections).

\section{Conclusion}

In the present paper we have verified (to leading order) the cutting rule for the polarization operator in general plane-wave background fields by an explicit calculation [see Eq. (\ref{eqn:paircreation_opticaltheorem}) and also \cite{ritus_1985,fradkin_quantum_1991}]. Furthermore, we derived a double-integral representation for the leading-order contribution to the polarization operator [see Eq. (\ref{eqn:polarizationoperator_realincoming_Wfunctions}) and also \cite{baier_interaction_1975,mitter_quantum_1975,dinu_vacuum_2014}]. By combining both results we obtained a compact double-integral representation for the total photon-decay probability within an arbitrarily shaped plane-wave laser pulse and for an arbitrary wave function of the incoming photon [see Eqs. (\ref{eqn:paircreation_decayratefinalA}), (\ref{eqn:paircreation_totalprobabilityfinal}) and (\ref{eqn:paircreation_linpolprobability})]. For a relativistically intense background field the result simplifies and one obtains the well-known average of the probability in a constant-crossed field over the laser pulse shape [see Eq. (\ref{eqn:paircreat_probabilityxilarge}) and also \cite{ritus_1985}]. \highlight{Our numerical calculations show that already for $\xi \lesssim 10$ it is sufficient to apply this so-called (local) constant-crossed field approximation (see Fig. \ref{fig:plot_comparison}). However, it underestimates the dependence of the probability on the CEP if the formation region becomes large, i.e. for $\xi \sim 1$ (see Fig. \ref{fig:plot_smallxicep}). In this regime the pair-production probability shows an oscillatory behaviour with respect to the CEP and the oscillation period depends on $\xi$.  Similarly, the CEP and other parameters of the pulse shape become important in the tunneling regime (see Fig. \ref{fig:plot_xichi} and Fig. \ref{fig:plot_cep}). Moreover, we have highlighted that the exponential decay of the photon wave function [see Eq. (\ref{eqn:exactphotonwavefunction})] must be taken into account if the total pair-creation probability becomes of order unity [see Eq. (\ref{eqn:photonsurvivalprobabilityB}) and Fig. \ref{fig:plot_xichicomparison}]. This is important for future experimental studies, as already with available laser technology ($\xi \sim 100$) a total pair-creation probability of the order of ten percent could be reached for a single GeV gamma photon (see Fig. \ref{fig:plot_photonenergy}).} 

\begin{acknowledgments}
S.M. would like to thank Andreas Fischer, Andreas Kaldun, Ben King, Anton W\"ollert and Enderalp Yakaboylu for fruitful discussions. Furthermore, he is grateful to the Studienstiftung des deutschen Volkes for financial support. All plots have been created with Matplotlib \cite{hunter_matplotlib_2007} and the GSL \cite{GSL} has been used for numerical calculations.
\end{acknowledgments}

\appendix

\section{Breit-Wheeler pair creation}
\label{sec:paircreationappendix}

To calculate the pair-creation probability, we describe the incoming photon by a wave packet
\begin{gather}
\label{eqn:paircreation_photonwavepacketstate}
\ket{\Phi,\eta} = \int \frac{d^3q'}{(2\pi)^3 \, 2\eps_{\spvec{q}'}} \, \eta(q') \ket{\Phi_{\spvec{q}'}},
\end{gather}
where $\eps_{\spvec{q}'} = \sqrt{\spvec{q}'^2}$. \highlight{This has the advantage, as shown in \cite{ilderton_scattering_2013,peskin_introduction_2008,ryder_quantum_1996}, of avoiding the appearance of squared delta functions and ambiguities in interpreting volume factors.} In Eq.~(\ref{eqn:paircreation_photonwavepacketstate}) we suppose that all components of the wave packet have the same polarization (the polarization indices are suppressed) and are on shell, i.e. $q'^2 = 0$. Furthermore, $\ket{\Phi_{\spvec{q}'}}$ denotes a momentum eigenstate of the photon field with relativistic normalization
\begin{gather}
\label{eqn:paircreation_photonstatecovariantnormalization}
\braket{\Phi_{\spvec{q}}|\Phi_{\spvec{q}'}}
= 
2\eps_{\spvec{q}} (2\pi)^3 \delta^3(\spvec{q}-\spvec{q}').
\end{gather}
The wave-packet state describes a single particle [$\braket{\Phi,\eta|\Phi,\eta} = 1$] if the envelope function obeys the covariant normalization condition
\begin{gather}
\label{eqn:paircreation_wavepacketenvelopenormalization}
\int \frac{d^3q'}{(2\pi)^3\, 2\eps_{\spvec{q}'}}   \, \abs{\eta(q')}^2 
= 1
\end{gather}
(this is assumed in the following). 

The probability that a single photon decays into an electron-positron pair inside a plane-wave background field is now given by 
\begin{gather}
\label{eqn:paircreation_decayprobabilityA}
\probsym
=
\sum_{\sigma,\sigma'} \int \frac{d^3p \, d^3p'}{(2\pi)^6 \, 2\eps_{\spvec{p}} 2\eps_{\spvec{p}'}} \,
\abs{\braket{\Phi_{\spvec{p},\sigma,\spvec{p}',\sigma'}|S|\Phi,\eta}}^2,
\end{gather}
where $\ket{\Phi_{\spvec{p},\sigma,\spvec{p}',\sigma'}}$ describes an electron and a positron with momenta $p^\mu=(\eps_{\spvec{p}},\spvec{p})$ and $p'^\mu=(\eps_{\spvec{p}'},\spvec{p}')$, respectively ($\sigma$,$\sigma'$ $\in [1,2]$ label the different spin states, $\eps_{\spvec{p}} = \sqrt{m^2+\spvec{p}^2}$, $\eps_{\spvec{p}'} = \sqrt{m^2+\spvec{p}'^2}$). Equation (\ref{eqn:paircreation_decayprobabilityA}) holds if the one-particle momentum eigenstates for the electron and the positron are relativistically normalized [see Eq.~(\ref{eqn:paircreation_photonstatecovariantnormalization})] \cite{maggiore_modern_2005},
\begin{gather}
\braket{\Phi_{\spvec{p},\sigma}|\Phi_{\spvec{p}',\sigma'}}
= 2\eps_{\spvec{p}} (2\pi)^3 \delta^3(\spvec{p}-\spvec{p}')\delta_{\sigma\sigma'},
\end{gather}
as the identity operator (in the one-particle subspace) is then given by
\begin{gather}
\one = \sum_{\sigma=1,2} \int \frac{d^3p}{(2\pi)^3 \, 2\eps_{\spvec{p}}} \ket{\Phi_{\spvec{p},\sigma}} \bra{\Phi_{\spvec{p},\sigma}}.
\end{gather}
In the following we drop the spin labels and write $\ket{\Phi_{\spvec{p},\spvec{p}'}} = \ket{\Phi_{\spvec{p},\sigma,\spvec{p}',\sigma'}}$ for simplicity.  Note that $\probsym$ is a probability (not a rate), as the duration of the process is naturally limited if the background field has only a finite extend.

Using Eq.~(\ref{eqn:paircreation_photonwavepacketstate}) we rewrite the squared matrix element in Eq.~(\ref{eqn:paircreation_decayprobabilityA}) as
\begin{multline}
\label{eqn:paircreation_squaredmatrixelementA}
\abs{\braket{\Phi_{\spvec{p},\spvec{p}'}|S|\Phi,\eta}}^2
=
\int \frac{d^3q_1 \, d^3q_2}{(2\pi)^6 \, 2\eps_{\spvec{q}_1} 2\eps_{\spvec{q}_2}} \eta(q_1) \eta^*(q_2)
\\\times\,
\mathfrak{M}(p,p';q_1) [\mathfrak{M}(p,p';q_2)]^*,
\end{multline}
where 
\begin{gather}
\I \mathfrak{M}(p,p';q)
=
\braket{\Phi_{\spvec{p},\spvec{p}'}|S|\Phi_{\spvec{q}}}
\end{gather}
[for simplicity we often suppress some of the labels, i.e. $\mathfrak{M}(p,\sigma,p',\sigma';q) = \mathfrak{M}(p,p';q) = \mathfrak{M}(q)$].

From now on we consider only plane-wave external fields (we use the same notation as in \cite{meuren_polarization_2013}, see also \cite{battesti_magnetic_2013,di_piazza_extremely_2012,ehlotzky_fundamental_2009,mourou_optics_2006,marklund_nonlinear_2006,dittrich_probingquantum_2000,fradkin_quantum_1991,ritus_1985,mitter_quantum_1975} for more details). In a plane-wave field the field tensor 
\begin{gather}
\label{eqn:fieldtensor}
F^{\mu\nu}(\phi) 
= 
f_1^{\mu\nu} \psi_1'(\phi)
+
f_2^{\mu\nu} \psi_2'(\phi),
\end{gather}
depends only on the laser phase $\phi = kx$, where $f_i^{\mu\nu} = k^\mu a_i^\nu - k^\nu a_i^\mu$ ($k^\mu$ characterizes the four-momentum of the background-field photons and the prime denotes the derivative with respect to the argument). This implies that the dressed vertex 
\begin{gather}
\label{eqn:sfqed_dressedvertex}
\Gamma^\rho(p,q,-p') = -\I e  \int d^4x \, e^{-\I qx}\, \bar{E}_{p,x} \gamma^\rho E_{-p',x},
\end{gather}
\begin{gather}
\begin{aligned}
E_{p,x} 
&= 
\lsb\one + \frac{e\s{k}\s{A}(kx)}{2\, kp}\rsb \, e^{iS_p(x)},
\\
\bar{E}_{p,x} 
&= 
\lsb\one + \frac{e\s{A}(kx)\s{k}}{2\, kp}\rsb \, e^{-iS_p(x)},
\end{aligned}
\end{gather}
with 
\begin{gather}
\label{eqn:sfqed_volkovphase}
S_p(x) 
=
- px - \int_{-\infty}^{kx} d\phi' \, \lsb \frac{e\, pA(\phi')}{\, kp} - \frac{e^2A^2(\phi')}{2\, kp} \rsb
\end{gather}
(see Sec. II.D of \cite{meuren_polarization_2013} for more details) contains three momentum-conserving delta functions
\begin{gather}
\label{eqn:dressedvertexfinal_momentumconservation}
\Gamma^\mu(p',q,p)
=
(2\pi)^3 \delta^{(\lminus,\lperp)}(p'-q-p) \, \mc{G}^\mu(p',q,p).
\end{gather}
Here we introduced light-cone coordinates \cite{dirac_forms_1949,neville_quantum_1971,mitter_quantum_1975}
\begin{gather}
\label{eqn:paircreation_lcc}
v^\lminus = vk,
\quad
v^\lplus = v\bar{k},
\quad
v^\lone	=  ve_1,
\quad
v^\ltwo	=  ve_2
\end{gather}
($v^\mu$ is an arbitrary four-vector, $\lone$ and $\ltwo$ are also summarized as $\lperp$), where we require that the four-vectors $k^\mu$, $\bar{k}^\mu$, $e_1^\mu$ and $e_2^\mu$ form a light-cone basis  [see Appendix C and Eq. (32) of \cite{meuren_polarization_2013}  for more details]. 

Thus, the $S$-matrix contains three overall momentum-conserving delta functions [see Eq.~(\ref{eqn:dressedvertexfinal_momentumconservation})] and  it is useful to define the reduced matrix element $\mc{M}$ by 
\begin{gather}
\label{eqn:paircreation_reducedmatrixelement}
\I \mathfrak{M}(p,p';q)
=
(2\pi)^3 \delta^{(\lminus,\lperp)}(p+p'-q) \, \I \mc{M}(p,p';q).
\end{gather}

To exploit the light-cone delta functions, we note the following relation between on-shell momentum integrals
\begin{multline}
\label{eqn:onshelllightconeintegrals}
\int \frac{d^3p}{(2\pi)^3} \frac{1}{2\eps_{\spvec{p}}} f(p)
=
\int \frac{d^4p}{(2\pi)^3} \delta(p^2-m^2)\theta(p^0) f(p)
\\=
\int \frac{dp^\lminus dp^\lperp}{(2\pi)^3} \frac{\theta(p^\lminus)}{2p^\lminus} f(p).
\end{multline}
Here $m$ is the particle mass ($p^2=m^2$), i.e. $p^0=\eps_{\spvec{p}}=\sqrt{m^2+\spvec{p}^2}$ in the first line and  $p^\lplus = \nfrac{(p^\lperp p^\lperp + m^2)}{(2p^\lminus)}$ in the last line. We note that $p^0 = \eps_{\spvec{p}}$ corresponds to $p^\lminus > 0$  and $p^0 = -\eps_{\spvec{p}}$ to $p^\lminus < 0$ ($p^\lminus = 0$ is only reached in the limit $\eps_{\spvec{p}} \to \infty$).

Hence, we can rewrite Eq.~(\ref{eqn:paircreation_squaredmatrixelementA}) as follows 
\begin{multline}
\label{eqn:paircreation_squaredmatrixelementfinal}
\abs{\braket{\Phi_{\spvec{p},\spvec{p}'}|S|\Phi,\eta}}^2
=
\int \frac{dq_1^\lminus dq_1^\lperp}{(2\pi)^3} \frac{\theta(q_1^\lminus)}{2q_1^\lminus} \, \abs{\eta(q_1)}^2 
\\ \times \,
\frac{1}{2q_1^\lminus} \abs{\mc{M}(p,p';q_1)}^2 \, 
(2\pi)^3 \delta^{(\lminus,\lperp)}(p+p'-q_1).
\end{multline}
Finally, the total probability for pair creation is given by [see Eq.~(\ref{eqn:paircreation_decayprobabilityA})]
\begin{subequations}
\label{eqn:paircreation_decayratefinalA}
\begin{gather}
\probsym
=
\int \frac{d^3q'}{(2\pi)^3\, 2\eps_{\spvec{q}'}} \abs{\eta(q')}^2 \, W(q'),
\end{gather}
where
\begin{multline}
\probsym(q)
=
\sum_{\text{spin}} \int \frac{d^3p \, d^3p'}{(2\pi)^6 \, 2\eps_{\spvec{p}} 2\eps_{\spvec{p}'}} \,
\frac{1}{2q^\lminus} \, \abs{\mc{M}(p,p';q)}^2 
\\ \times \,
(2\pi)^3 \delta^{(\lminus,\lperp)}(p+p'-q).
\end{multline}
\end{subequations}

Using the Feynman rules for QED with plane-wave background fields (see \cite{meuren_polarization_2013} for details), we obtain the following matrix element for the diagram in Fig.~\ref{fig:pair}:
\begin{gather}
\label{eqn:paircreation_matrixelement}
\mathfrak{M}(p,\sigma,p',\sigma';q)
=
\eps_\mu \, \bar{u}_{p,\sigma} \Gamma^\mu(p,q,-p') v_{p',\sigma'},
\end{gather}
where $\eps_\mu$ is the polarization four-vector of the incoming photon ($\eps q = 0$, $\eps^\mu \eps^*_\mu =-1$). The four-spinors of the electron ($u_{p,\sigma}$) and the positron ($v_{p',\sigma'}$) obey \cite{landau_quantum_1981,peskin_introduction_2008}
\begin{gather}
(\s{p}-m)u_{p,\sigma} = 0,
\quad
(\s{p}'+m)v_{p',\sigma'} = 0.
\end{gather}

\highlight{We point out that the matrix element in Eq.~(\ref{eqn:paircreation_matrixelement}) represents only the leading-order contribution to the pair-creation process. Furthermore, $\probsym$ can only be interpreted as the probability for pair production as long as it is small. In general, it represents the decay exponent of the exact photon wave function (see Sec. \ref{sec:exactphotonwavefunction} for more details).}

\section{Cutting rules for the polarization operator}
\label{sec:cuttingrulesappendix}

\highlight{In this Appendix we will explicitly derive the optical theorem for pair-creation (to leading order) in the presence of a plane-wave background field (see also \cite{ritus_vacuum_1972,baier_interaction_1975,fradkin_quantum_1991,di_piazza_barrier_2009,milstein_polarization-operator_2006,dinu_vacuum_2014} and e.g. \cite{landau_analytic_1959,cutkosky_singularities_1960,peskin_introduction_2008,srednicki_quantum_2007} for the corresponding proof in vacuum QED).} To this end we consider the squared matrix element [see Eq.~(\ref{eqn:paircreation_matrixelement})], which appears in Eq.~(\ref{eqn:paircreation_squaredmatrixelementA})
\begin{multline}
\label{eqn:paircreation_matrixelementsquared}
\mathfrak{M}(p,\sigma,p',\sigma';q_1) [\mathfrak{M}(p,\sigma,p',\sigma';q_2)]^*
\\=
\eps_\mu \eps^*_\nu \, \tr \rho^u_{p,\sigma} \Gamma^\mu(p,q_1,-p') \rho^v_{p',\sigma'}  \overline{\Gamma}^\nu(p,q_2,-p')
\end{multline}
[see Eq.~(\ref{eqn:fundamentalgammamatricesbared}) for the bar notation used]. Here we have introduced the density matrices
\begin{gather}
\rho^u_{p,\sigma} = u_{p,\sigma} \bar{u}_{p,\sigma},
\quad
\rho^v_{p',\sigma'} = v_{p',\sigma'} \bar{v}_{p',\sigma'}.
\end{gather}
To obtain the total pair-creation probability we have to sum/integrate over final spins and momenta [see Eq.~(\ref{eqn:paircreation_decayprobabilityA})]
\begin{gather}
\label{eqn:paircreation_squaredmatrixelementintegrated}
\sum_{\text{spin}} \int \frac{d^3p \, d^3p'}{(2\pi)^6 \, 2\eps_{\spvec{p}} 2\eps_{\spvec{p}'}} \, \mathfrak{M}(q_1) [\mathfrak{M}(q_2)]^*.
\end{gather}
The sum over different spin states yields \cite{landau_quantum_1981,peskin_introduction_2008}
\begin{gather}
\sum_{\sigma=1,2} \rho^u_{p,\sigma} = \s{p} + m,
\quad
\sum_{\sigma'=1,2} \rho^v_{p',\sigma'} = \s{p}' - m.
\end{gather}
Thus, we see that Eq.~(\ref{eqn:paircreation_squaredmatrixelementintegrated}) resembles the leading-order contribution to the polarization operator (see Fig.~\ref{fig:polop} and Appendix \ref{sec:poloppaperresult})
\begin{multline}
\label{eqn:polarizationoperator}
\I \mc{P}^{\mu\nu}(q_1,q_2)
\\=
\int  \frac{d^4p_1\, d^4p_2}{(2\pi)^8} 
\frac{\tr [\cdots]^{\mu\nu}}{(p_1^2-m^2+\I0)(p_2^2-m^2+\I0)},
\end{multline}
where
\begin{multline}
\label{eqn:paircreation_polarizationoperatortrace}
\tr [\cdots]^{\mu\nu}
=
\tr 
\, 
\Gamma^\mu(p_2,q_1,p_1)  (\s{p}_1 + m) 
\\ \times \Gamma^\nu(p_1,-q_2,p_2) (\s{p}_2 + m).
\end{multline}
To match the two expressions even further, we introduce two more integrations in $p^0$ and $p'^0$ in Eq.~(\ref{eqn:paircreation_squaredmatrixelementintegrated}) together with appropriate delta and step functions to bring the momenta on shell [see Eq.~(\ref{eqn:onshelllightconeintegrals})]. After applying the identity
\begin{gather}
\label{eqn:sfqed_dressedvertexbared}
\overline{\Gamma}^\rho(p',q,p) 
= 
-\Gamma^\rho(p,-q,p'),
\end{gather}
which follows from
\begin{gather}
\label{eqn:fundamentalgammamatricesbared}
\begin{gathered}
\overline{\one}
=
\one,
\quad
\overline{\gamma^{5}} 
=
- \gamma^5,
\quad
\overline{\gamma^\mu}
=
\gamma^\mu,
\\
\overline{\lb \I\gamma^\mu \gamma^5 \rb} 
=
- \I\gamma^\mu \gamma^5,
\quad
\overline{\lb \I\sigma^{\mu\nu} \rb} 
=
\I\sigma^{\mu\nu},
\end{gathered}
\end{gather}
using the cyclic property of the trace and the change of variables $p^\mu \to p_2^\mu$, $p'^\mu \to -p_1^\mu$ we obtain
\begin{multline}
\label{eqn:paircreation_matrixelementsummed2}
\sum_{\text{spin}} \int \frac{d^3p \, d^3p'}{(2\pi)^6 \, 2\eps_{\spvec{p}} 2\eps_{\spvec{p}'}} \, \mathfrak{M}(q_1) [\mathfrak{M}(q_2)]^*
\\=
\int \frac{d^4p_1 \, d^4p_2}{(2\pi)^6} \,
\delta(p^2_1-m^2) \delta(p^2_2-m^2) \theta(-p^0_1) \theta(p_2^0)
\\
\times \eps_\mu \eps^*_\nu \tr [\cdots]^{\mu\nu}.
\end{multline}

To prove the optical theorem we have to relate the imaginary part of the forward photon scattering amplitude to the total pair-creation probability. Therefore, we extract the nonsingular part of the polarization operator by defining
\begin{gather}
\label{eqn:polopnonsingularpart}
\mc{P}^{\mu\nu}(q_1,q_2)
=
(2\pi)^3 \delta^{(\lminus,\lperp)}(q_1-q_2) \, \Pi^{\mu\nu}(q_1,q_2)
\end{gather}
and consider $\Im \, [\eps_\mu \eps^*_\nu \Pi^{\mu\nu}(q,q)]$. We point out that the contracted trace $\eps_\mu \eps^*_\nu \tr [\cdots]^{\mu\nu}$ [see Eq.~(\ref{eqn:paircreation_polarizationoperatortrace})] is purely real if evaluated at $q_1^\mu = q_2^\mu = q^\mu$ (strictly speaking, after the singular part is factorized out). This can be deduced from
\begin{gather}
\label{eqn:paircreation_polarizationoperatortracecc}
(\tr [\cdots]^{\mu\nu})^* = \tr [\cdots]^{\nu\mu}(q_1 \leftrightarrow q_2)
\end{gather}
[note that $(\tr M)^* = \tr M^\dagger = \tr \bar{M}$]. Using the Sokhotski-Plemelj identity \cite{srednicki_quantum_2007,merzbacher_quantum_1998}
\begin{gather}
\label{eqn:paircreation_sokhotskiplemelj}
\frac{1}{p^2- m^2 + \I0}
=
\PP \frac{1}{p^2 - m^2}
- \I \pi \delta \lb p^2 - m^2\rb,
\end{gather}
we obtain the symbolic relation [see Eq.~(\ref{eqn:polarizationoperator})]
\begin{gather}
\Im \, [\eps_\mu \eps^*_\nu \Pi^{\mu\nu}(q,q)]
= - \Re \, [\eps_\mu \eps^*_\nu \I \Pi^{\mu\nu}(q,q)] \sim  \pi^2 \delta \delta - \PP\PP.
\end{gather}
It is shown in Appendix \ref{sec:polestructurevolkov} that the two principle value integrals are related to the on-shell contribution. Symbolically, the result can be written as
\begin{gather}
\label{eqn:polopcut_principlevsdelta}
\PP\PP = \sign(p_1^\lminus)\sign(p_2^\lminus) \pi^2 \delta\delta,
\end{gather}
implying
\begin{gather}
\Im \, [\eps_\mu \eps^*_\nu \Pi^{\mu\nu}(q,q)]   \sim [1-\sign(p_1^\lminus)\sign(p_2^\lminus)] \pi^2 \delta \delta.
\end{gather}

On the other hand, the momentum-conserving delta function $\delta^{(\lminus)}(p_2-q_1-p_1)$ contained in the vertices in Eq. (\ref{eqn:paircreation_matrixelementsummed2}) ensures that only the region $p_2^\lminus - p_1^\lminus > 0$ contributes to the integral (assuming $q_i^\lminus > 0$, i.e. we exclude the trivial case of a photon which is copropagating with the laser). Thus, in Eq.~(\ref{eqn:paircreation_matrixelementsummed2}) we can apply the replacement
\begin{gather}
2\theta(-p_1^0)\theta(p_2^0) 
\longleftrightarrow 
[1 - \sign(p_1^\lminus)\sign(p_2^\lminus)]. 
\end{gather}
Finally, we obtain
\begin{multline}
\label{eqn:paircreation_imforwardscatteringamplitude}
2 \Im \, [\eps_\mu \eps^*_\nu \, \Pi^{\mu\nu}(q,q)]
=
\eps_\mu \eps^*_\nu  \, \int \frac{d^3p \, d^3p'}{(2\pi)^6 \, 2\eps_{\spvec{p}} 2\eps_{\spvec{p}'}} 
\\\times
(2\pi)^3 \delta^{(\lminus,\lperp)}(p+p'-q)
\, \tr (\s{p} + m) \mc{G}^\mu(p,q,-p') 
\\\times
(\s{p}' - m)  \overline{\mc{G}}^\nu(p,q,-p')
\end{multline}
(for $q^\lminus > 0$), where $\mc{G}^\mu$ denotes the nonsingular part of the dressed vertex [see Eq.~(\ref{eqn:dressedvertexfinal_momentumconservation})]. By combining everything, we obtain the following relation between the total nonlinear Breit-Wheeler pair-creation probability $W$ and the imaginary part of the photon forward-scattering amplitude (see also \cite{di_piazza_barrier_2009,milstein_polarization-operator_2006,baier_interaction_1975})
\begin{subequations}
\label{eqn:paircreation_opticaltheorem}
\begin{gather}
\probsym(q)
=
\frac{1}{kq} \, \Im \, [\eps_\mu \eps^*_\nu \, \Pi^{\mu\nu}(q,q)]
\end{gather}
($q^2=0$) and [see Eq.~(\ref{eqn:paircreation_decayratefinalA})]
\begin{gather}
\probsym
=
\int \frac{d^3q'}{(2\pi)^3\, 2\eps_{\spvec{q}'}} \abs{\eta(q')}^2 \, W(q')
\end{gather}
\end{subequations}
[$W \approx W(q)$ if the wave packet of the incoming photon is sharply peaked around $q'^\mu=q^\mu$, see Eq.~(\ref{eqn:paircreation_wavepacketenvelopenormalization})].

\section{Pole structure of the Volkov propagator}
\label{sec:polestructurevolkov}

To prove Eq.~(\ref{eqn:polopcut_principlevsdelta}) we have to investigate the pole structure of the the Volkov propagator
\begin{gather}
\label{eqn:dressedpropagator}
\I G(x,y) 
= 
\I \int \frac{d^4p}{(2\pi)^4} E_{p,x} \frac{\s{p} + m}{p^2 - m^2 + \I0} \bar{E}_{p,y},
\end{gather}
which describes the propagation of a fermion from $y$ to $x$ (and correspondingly the propagation of an antifermion from $x$ to $y$), taking the plane-wave background field into account exactly. This is most conveniently carried out in light-cone coordinates, where the integral in $dp^\lplus$ has a simple structure \cite{ritus_radiative_1972}, as the phase of the propagator depends on $p^\lplus$ only via [see Eq.~(\ref{eqn:sfqed_volkovphase})]
\begin{gather}
\exp \lsb - \I p^\lplus (x^\lminus-y^\lminus)\rsb
\end{gather}
($A^\lminus=k^\lminus =0$). For $p^\lminus \neq 0$ we can evaluate the integral in $p^\lplus$ using the residue theorem \cite{ablowitz_complex_2003}. 

\highlight{In general, the point $p^\lminus=0$ (which corresponds to the so-called light-cone zero mode) must be treated with care (for more details about light-cone quantization and the light-cone zero mode see e.g. \cite{dirac_forms_1949,brodsky_quantum_1998,heinzl_light-cone_2001}). As long as no singularities [e.g. a delta function $\delta(p^\lminus)$] are encountered, a single point can always be excluded from the integration range. In the absence of external fields such a delta function appears in QED only in diagrams without external legs and for all other diagrams the light-cone zero mode can be ignored (see Sec. II.C in \cite{chang_feynman_1969}).  Like in vacuum QED, the light-cone zero mode does not contribute to the leading-order diagram for the polarization operator in a plane-wave background field if the incoming photon is on shell and does not propagate collinearly with the laser field. This can be seen explicitly from the final expression of the field-dependent part of the polarization operator given in \cite{meuren_polarization_2013} [see Eqs. (92)-(97) there]. In fact, the integrand of the polarization operator vanishes at the points $\tau=0$ and $v=\pm 1$, corresponding to vanishing values of at least one of the proper-time variables $t$ and $s$. Thus, the delta function used to take the $p^\lminus$-integral [Eq. (55) in \cite{meuren_polarization_2013}] implies that $p^\lminus \neq 0$ as long as $kq = q^\lminus \neq 0$. In conclusion, for the discussion of the optical theorem we can ignore subtleties arising from the light-cone zero mode and assume that $p^\lminus \neq 0$ in the following.}

To take the integral in $p^\lplus$ we have to close the contour in the lower complex plane if $x^\lminus-y^\lminus > 0$ and in the upper complex plane if $x^\lminus-y^\lminus < 0$. The pole of
\begin{gather}
\frac{1}{p^2 - m^2 + \I0}
=
\frac{1}{2p^\lplus p^\lminus - p^\lperp p^\lperp - m^2 + \I0}
\end{gather}
is located at
\begin{gather}
p^\lplus = \frac{p^\lperp p^\lperp + m^2 - \I0}{2p^\lminus},
\end{gather}
i.e. in the lower complex plane for $p^\lminus>0$ and in the upper complex plane for $p^\lminus<0$, in agreement with the Feynman boundary condition. 

Following \cite{srednicki_quantum_2007}, we have to consider also the retarded and advanced propagators, defined by the pole prescriptions
\begin{gather}
\frac{1}{p^2 - m^2 + \sign(p^\lminus) \I0},
\quad
\frac{1}{p^2 - m^2 - \sign(p^\lminus) \I0},
\end{gather}
respectively. The pole of the former is always located in the lower, the pole of the latter always in the upper complex plane. Correspondingly, the propagators vanish for $x^\lminus < y^\lminus$ and $x^\lminus > y^\lminus$, respectively. 

The polarization operator diagram [see Eq.~(\ref{eqn:polarizationoperator})] contains both $G(x,y)$ and $G(y,x)$ or in other words the phase factor contains
\begin{gather}
\exp \lsb - \I p_1^\lplus (x^\lminus-y^\lminus)\rsb \exp \lsb \I p_2^\lplus (x^\lminus-y^\lminus)\rsb.
\end{gather}
Correspondingly, the contour integrals in $p_1^\lplus$ and $p_2^\lplus$ must be closed differently. If both propagators of the polarization operator are either replaced by advanced or by retarded propagators, such that for $x^\lminus-y^\lminus \gtrless 0$ one propagator is always zero,  the contribution of the total diagram vanishes. Using the relation [see Eq.~(\ref{eqn:paircreation_sokhotskiplemelj})]
\begin{multline}
\frac{1}{p^2- m^2 \pm \sign(p^\lminus) \I0}
=
\PP \frac{1}{p^2 - m^2}
\\
\mp \I \sign(p^\lminus) \pi \delta \lb p^2 - m^2\rb,
\end{multline}
we can now prove the identity [see Eq.~(\ref{eqn:polopcut_principlevsdelta})]
\begin{gather}
\PP\PP = \sign(p_1^\lminus)\sign(p_2^\lminus) \pi^2\delta\delta
\end{gather}
for $\Im \, [\eps_\mu \eps^*_\nu \Pi^{\mu\nu}(q,q)]$.

\section{Polarization operator}
\label{sec:poloppaperresult}

For plane-wave background fields the polarization operator was first considered in \cite{baier_interaction_1975,becker_vacuum_1975} [see Eq.~(\ref{eqn:polarizationoperator}), Fig. \ref{fig:polop} and also \cite{meuren_polarization_2013,dinu_vacuum_2014,dinu_photon_2014,gies_laser_2014} for a recent discussion]. Starting from Eq. (92) in \cite{meuren_polarization_2013} [there the notation $\I\mc{P}^{\mu\nu}(q_1,q_2) = T^{\mu\nu}(q_1,q_2)$ is used], a double-integral representation for the polarization operator can be derived \highlight{(for other double-integral representations see \cite{dinu_vacuum_2014,becker_vacuum_1975})}. To this end we apply the change of variables from $\tau = \mu w = \nfrac{\varrho w}{kq}$ to $\varrho = \mu kq$ [$\mu = \frac14 \tau (1-v^2)$, $\frac{1}{w} = \frac14(1-v^2)$] and from $v$ to $w$
\begin{multline}
\label{eqn:polop_integralchangedtophysicalparameters}
\int_{-1}^{+1} dv \int_0^\infty \frac{d\tau}{\tau} \, 
\int_{-\infty}^{+\infty} dz^\lminus
\\=
\int_{4}^{\infty} dw \, \frac{4}{w\sqrt{w(w-4)}} \int_0^{\sigma\infty} \frac{d\varrho}{\varrho} \,  \int_{-\infty}^{+\infty} dz^\lminus,
\end{multline}
where $\sigma = \sign(kq)$ [we assume that the integrand is an even function of $v$, see Eq.~(\ref{eqn:paircreation_vtowint})]. As in \cite{meuren_polarization_2013} we simply write $q^\mu$ if $q_1^\mu$ and $q_2^\mu$ can be used interchangeably due to the momentum-conserving delta functions.

The new variables have a clear physical meaning, as the phases of the creation and the annihilation vertex are given by $kx = kz - \varrho$ and $ky = kz + \varrho$, respectively, and the variable $w$ is related to the momenta $p^\mu$ and $p'^\mu$ of the created electron and positron, respectively, by $w = \nfrac{(kq)^2}{(kp kp')}$ [the momenta $p^\mu$ and $p'^\mu$ here differ from those denoted by the same symbols in \cite{meuren_polarization_2013}, see Eq.~(\ref{eqn:polarizationoperator})].

In terms of the new variables the phases [see Eq. (93) and Eq. (95) in \cite{meuren_polarization_2013}] can be written as 
\begin{gather}
\label{eqn:polop_phases}
\begin{aligned}
\Phi
&=
(q_2^\lplus -q_1^\lplus) z^\lminus 
+ 
\varrho (\nfrac{q_1q_2}{kq}) 
- 
w \, (\nfrac{m^2}{kq}) \, \varrho,
\\
\Phi_1
&=
(q_2^\lplus -q_1^\lplus) z^\lminus 
+ 
\varrho \, (\nfrac{q_1q_2}{kq}) 
- 
w \, (\nfrac{m^2}{kq}) \, \mc{D}(\varrho,kz),
\end{aligned}
\end{gather}
where $\Phi_1 = \Phi + \tau\beta$ and we defined [see Eq. (96) in \cite{meuren_polarization_2013}]
\begin{gather}
\begin{gathered}
\label{eqn:polop_massdressing}
\mc{D}(\varrho,kz)
= 
\varrho \, \Big[1 + \sum_{i=1,2} \xi_i^2 \big(J_i-I_i^2 \big) \Big],
\\
I_i 
=
\frac{1}{2\varrho} \int_{kz - \varrho}^{kz + \varrho} d\phi \, \psi_i(\phi),
\quad
J_i 
=
\frac{1}{2\varrho} \int_{kz - \varrho}^{kz + \varrho} d\phi \, \psi^2_i(\phi).
\end{gathered}
\end{gather}
Thus, after the change of variables given in Eq.~(\ref{eqn:polop_integralchangedtophysicalparameters}), the phases have a very simple dependence on $w$ and the integral in $w$ can be calculated analytically.

\begin{figure}[t]
\centering
\includegraphics{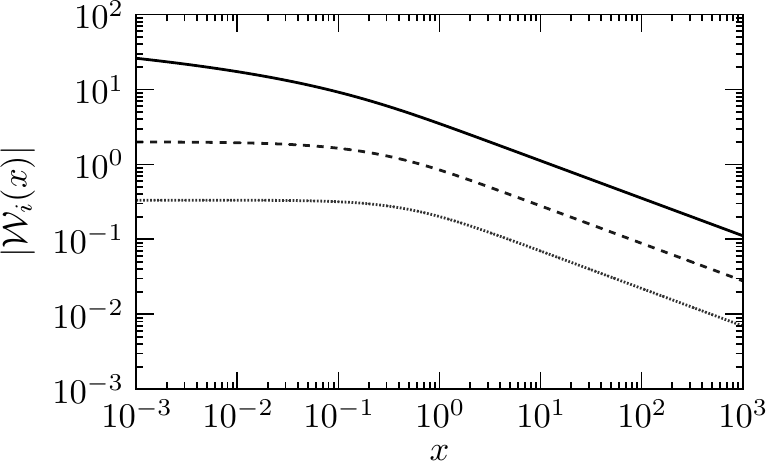}
\caption{\label{fig:wifunctions}The function $\mc{W}_0$ (solid line) has a logarithmic singularity at the origin, $\mc{W}_1$ (dashed line) and $\mc{W}_2$ (dotted line) are regular. For large values they behave as $\sim x^{-\nfrac12}$.}
\end{figure}

To this end we define the functions $\mc{W}_l(x)$ ($l=0,1,2$, $x\geq 0$, see Fig. \ref{fig:wifunctions})
\begin{gather}
\label{eqn:polop_dwfouriertransform}
\int_{4}^{\infty} dw \, \frac{4}{w^l \sqrt{w(w-4)}} \,  e^{-\I w x}
=
e^{-\I 4x} \, \mc{W}_l(x),
\end{gather}
which are non-oscillatory and scale asymptotically as
\begin{gather}
\label{eqn:polop_Wileadingorderlargeargument}
\mc{W}_l(x)
\sim
-\frac{2\sqrt{\pi}\I}{4^{l}} \, e^{\I \nfrac{\pi}{4}} \frac{1}{\sqrt{x}}
\end{gather}
[note that $\mc{W}_0(x)$ has a logarithmic singularity at $x=0$]. Using the following integral representation for the Hankel function \cite{gradshteyn_table_2007}
\begin{gather}
\label{eqn:polop_hankelintegralrep}
\Hankel_{\nu}^{(2)}(z) = \I \frac{2}{\pi} e^{\nfrac{\nu \pi \I}{2}} \int_0^\infty dt \, e^{-\I z \cosh t} \cosh (\nu t)
\end{gather}
(valid for $-1<\Re \nu<1$, $z>0$), we finally obtain
\begin{gather}
\label{eqn:polop_doubleintrep_Wfunctions}
\begin{aligned}
\mc{W}_0(x) &=  (-2\pi \I) \, e^{\I 2x} \, \Hankel_0^{(2)}(2x),
\\
\mc{W}_1(x) &=  (-2\pi x) \, e^{\I 2x} \, \big[ \Hankel_0^{(2)}(2x)  + \I \Hankel_1^{(2)}(2x) \big],
\\
\mc{W}_2(x) &= \frac{\pi x}{3} \, e^{\I 2x} \, \big[ 4\I x \Hankel_0^{(2)}(2x) - (4x+\I) \Hankel_1^{(2)}(2x) \big].
\end{aligned}
\end{gather}

The representation for the polarization operator given in \cite{meuren_polarization_2013} [see Eq. (92) there] depends on the external momenta via the scalar $q_1q_2$. However, for real incoming or outgoing photons it is more convenient to use a representation which depends only on $q_1^2$ or $q_2^2$. To obtain such a representation, we use the three momentum-conserving delta functions and write
\begin{gather}
\label{eqn:polop_momentumconservationwithn}
\begin{gathered}
q_2^\mu = q_1^\mu + n k^\mu,
\quad
n = q_2^\lplus -q_1^\lplus,
\\
q_1q_2 = q_1^2 + n kq = q_2^2 - n kq,
\end{gathered}
\end{gather}
where $n$  denotes the amount of four-momentum $k^\mu$ exchanged with the background field ($n>0$ corresponds to absorption, $n<0$ to emission, $n$ is in general not an integer). Thus, the integral in $z^\lminus$ represents a Fourier transform that determines the probability amplitude to absorb $nk^\mu$ four-momentum from the background field. Correspondingly, the phases [see Eq.~(\ref{eqn:polop_phases})] can now be rewritten using
\begin{multline}
\label{eqn:polop_phaseidentity}
(q_2^\lplus -q_1^\lplus) z^\lminus 
+ 
\varrho \frac{q_1q_2}{kq}  
\\=
n kz + \varrho \frac{q_1q_2}{kq}
=
n ky + \varrho \frac{q_1^2}{kq}
=
n kx + \varrho \frac{q_2^2}{kq}.
\end{multline}
Thus, by changing the integration variable from $z^\lminus$ to either $x^\lminus$ (real outgoing photon) or $y^\lminus$ (real incoming photon) the phase of the polarization operator simplifies in these cases. Depending on this choice one of the following representations is convenient [see Eq.~(\ref{eqn:polop_massdressing})]
\begin{gather}
\label{eqn:polop_IiJiforkxkykz}
\begin{aligned}
I_i 
&=
\int_{0}^{1} d\lambda\, \psi_i(ky - 2\varrho \lambda)
=
\int_{0}^{1} d\lambda \, \psi_i(kx + 2\varrho \lambda),
\\
J_i 
&=
\int_{0}^{1} d\lambda\, \psi^2_i(ky - 2\varrho \lambda)
=
\int_{0}^{1} d\lambda \, \psi^2_i(kx + 2\varrho \lambda).
\end{aligned}
\end{gather}

Similarly, we can rewrite the preexponent using the following identity
\begin{multline}
\label{eqn:polop_nkqrewritten}
\frac{n kq}{2} \, \int_{-\infty}^{+\infty} dz^\lminus e^{\I \Phi} \lb e^{\I\tau\beta} - 1\rb
\\
\begin{aligned}
&=
(-\I) \frac{kq}{2} \, \int_{-\infty}^{+\infty} dz^\lminus \lb e^{\I\tau\beta} - 1\rb  \frac{\del}{\del z^\lminus} e^{\I \Phi}
\\&=
2m^2 \frac{\tau}{4\mu} \int_{-\infty}^{+\infty} dz^\lminus  e^{\I \Phi} e^{\I\tau\beta}  \sum_{i=1,2} \xi_i^2 \lb Y_i - Z_i \rb,
\end{aligned}
\end{multline}
where
\begin{gather}
Y_i 
= 
[I_i-\psi_i(ky)] \, [\psi_i(kx) - \psi_i(ky)].
\end{gather}
To prove Eq.~(\ref{eqn:polop_nkqrewritten}) we used integration by parts and
\begin{gather}
\begin{aligned}
\frac{\del I_i(\varrho,kz)}{\del z^\lminus} &= - \frac{1}{2\varrho} \lsb \psi_i(kz - \varrho) - \psi_i(kz+\varrho) \rsb,\\
\frac{\del J_i(\varrho,kz)}{\del z^\lminus} &= - \frac{1}{2\varrho} \lsb \psi^2_i(kz - \varrho) - \psi^2_i(kz+\varrho) \rsb.
\end{aligned}
\end{gather}
Furthermore, it is useful to define $V_i = 2Z_i-Y_i$.

By applying the above relations to the symmetric representation given in Eq. (92) of \cite{meuren_polarization_2013}, we immediately obtain the representation given in Eq. (109) of \cite{meuren_polarization_2013}, which is equivalent to the one in \cite{baier_interaction_1975}. Moreover, using Eq. (\ref{eqn:polop_dwfouriertransform}) we obtain for the field-dependent part of the polarization operator inside a plane-wave background field the following double-integral representation
\begin{multline}
\label{eqn:polarizationoperator_realincoming_Wfunctions}
\I\mc{P}^{\mu\nu}(q_1,q_2)
-
\I\mc{P}^{\mu\nu}_{\Ftilde=0}(q_1,q_2)
= 
- \I (2\pi)^3 \, \delta^{(\lminus,\lperp)}(q_1-q_2) \,
\\
\times \, \frac{\alpha}{2\pi} \, \int_0^{\sigma\infty} \frac{d\varrho}{\varrho} \,  \int_{-\infty}^{+\infty} dy^\lminus
\big[
P_{12} \Lambda_1^\mu \Lambda_2^\nu
+
P_{21} \Lambda_2^\mu \Lambda_1^\nu
\\+
P_{11} \Lambda_1^\mu \Lambda_1^\nu
+
P_{22} \Lambda_2^\mu \Lambda_2^\nu
+
P_{Q} \Q_1^\mu \Q_2^\nu
\big],
\end{multline}
$\sigma = \sign(kq)$, where the coefficients are now given by
\begin{align}
P_{12} &= \frac{m^2 \xi_1 \xi_2}{2} \big\{  \mc{W}_0(x_1) X_{12} \nonumber\displaybreak[0]\\\nonumber
&\phantom{= }+ [4\mc{W}_1(x_1)-\mc{W}_0(x_1)] X_{21}  \big\} e^{i\widetilde{\Phi}_1},\nonumber\displaybreak[0]\\\nonumber
P_{21} &= \frac{m^2 \xi_1 \xi_2}{2} \big\{  \mc{W}_0(x_1) X_{21} \nonumber\displaybreak[0]\\\nonumber
&\phantom{= }+ [4\mc{W}_1(x_1)-\mc{W}_0(x_1)] X_{12}  \big\} e^{i\widetilde{\Phi}_1},\nonumber\displaybreak[0]\\\nonumber
P_{11} &= - m^2\bigg[ \frac{i}{\varrho} \frac{kq}{m^2} \mc{W}_2(x_1) + \frac{q_1^2}{2m^2} \mc{W}_1(x_1) \bigg] e^{i\widetilde{\Phi}_1}  \\\nonumber
&\phantom{= }+ m^2\bigg[ \frac{i}{\varrho} \frac{kq}{m^2} \mc{W}_2(x_0) + \frac{q_1^2}{2m^2} \mc{W}_1(x_0) \bigg] e^{i\widetilde{\Phi}_0}  \\\nonumber
&\phantom{= }+ m^2 \bigg[ \frac{1}{2} \lb \xi_1^2 V_1 + \xi_2^2 V_2 \rb \mc{W}_0(x_1) \\\nonumber
&\phantom{= }+ 2 \xi_1^2 X_{11} \mc{W}_1(x_1) \bigg] e^{i\widetilde{\Phi}_1},
\nonumber\displaybreak[0]\\\nonumber
P_{22} &= - m^2\bigg[ \frac{i}{\varrho} \frac{kq}{m^2} \mc{W}_2(x_1) + \frac{q_1^2}{2m^2} \mc{W}_1(x_1) \bigg] e^{i\widetilde{\Phi}_1}  \\\nonumber
&\phantom{= }+ m^2\bigg[ \frac{i}{\varrho} \frac{kq}{m^2} \mc{W}_2(x_0) + \frac{q_1^2}{2m^2} \mc{W}_1(x_0) \bigg] e^{i\widetilde{\Phi}_0}  \\\nonumber
&\phantom{= }+ m^2 \bigg[ \frac{1}{2} \lb \xi_1^2 V_1 + \xi_2^2 V_2 \rb \mc{W}_0(x_1) \\\nonumber
&\phantom{= }+ 2 \xi_2^2 X_{22} \mc{W}_1(x_1) \bigg] e^{i\widetilde{\Phi}_1},
\nonumber\displaybreak[0]\\
P_{Q} &= - 2  \lsb \mc{W}_2(x_1) e^{i\widetilde{\Phi}_1} - \mc{W}_2(x_0) e^{i\widetilde{\Phi}_0} \rsb
\label{eqn:polarizationoperator_realincoming_coefficients}
\end{align}
with the phases [see Eq.~(\ref{eqn:polop_phases})]
\begin{gather}
\label{eqn:polop_doubleint_realincoming_phases}
\begin{aligned}
\widetilde{\Phi}_0
&=
(q_2^\lplus -q_1^\lplus) y^\lminus 
+ 
\varrho (\nfrac{q_1^2}{kq}) 
- 
4 x_0,
\\
\widetilde{\Phi}_1
&=
(q_2^\lplus -q_1^\lplus) y^\lminus 
+ 
\varrho \, (\nfrac{q_1^2}{kq}) 
- 
4 x_1.
\end{aligned}
\end{gather}
Here we have introduced
\begin{gather}
\label{eqn:x0x1def}
x_0 = (\nfrac{m^2}{kq}) \, \varrho,
\quad
x_1 = (\nfrac{m^2}{kq}) \, \mc{D}(\varrho, ky),
\end{gather}
where [see Eq.~(\ref{eqn:polop_massdressing}) and Eq.~(\ref{eqn:polop_IiJiforkxkykz})]
\begin{gather}
\begin{gathered}
\mc{D}(\varrho, ky)
= 
\varrho \, \Big[1 + \sum_{i=1,2} \xi_i^2 \big( J_i-I_i^2 \big) \Big],
\\
I_i 
=
\int_{0}^{1} d\lambda\, \psi_i(ky - 2\varrho \lambda),
\quad
J_i 
=
\int_{0}^{1} d\lambda\, \psi^2_i(ky - 2\varrho \lambda).
\end{gathered}
\end{gather}
Furthermore, [see Eq. (97) in \cite{meuren_polarization_2013}]
\begin{gather}
\label{eqn:polop_XVofky}
\begin{aligned}
X_{ij} 
&= 
[I_i-\psi_i(ky)] \, [I_j - \psi_j(ky - 2\varrho)],\\
V_i 
&= 
\lsb I_i - \psi_i(ky-2\varrho) \rsb \lsb \psi_i(ky) - \psi_i(ky-2\varrho) \rsb.
\end{aligned}
\end{gather}

Having taken the $w$-integral analytically, we are left with the integrals in $y^\lminus$ and $\varrho$. To evaluate these integrals, the precise shape of the background field has to be known. It is therefore reasonable to use numerical methods (see appendix \ref{sec:numericalintegration}).

\section{Dressed mass}
\label{sec:polopmassdressing}

It is well known that inside a linearly polarized, monochromatic field the square of the dressed electron (positron) mass is given by \cite{ritus_1985}
\begin{gather}
m_*^2 = m^2 (1 + \nfrac{\xi^2}{2}),
\end{gather}
which corresponds to the square of the average (classical) electron four-momentum. This definition of the dressed mass may be generalized to an arbitrary plane-wave field by noting that the classical four-momentum of an electron (charge $e$ and mass $m$) is given by \cite{di_piazza_extremely_2012,sarachik_classical_1970}
\begin{gather}
\label{eqn:polop_momentumfourvectorsolution}
P^\mu(\phi)
=
P_{0}^\mu + \frac{e \Ftilde^{\mu}_{\phantom{\mu}\nu}(\phi,\phi_0)  P^{\nu}_{0}}{kP_{0}} 
+
\frac{e^2 \Ftilde^{2\mu}_{\phantom{2\mu}\nu}(\phi,\phi_0) P^{\nu}_{0}}{2(kP_{0})^2},
\end{gather}
where $P_{0}^\mu = P_{}^\mu(\phi_0)$ and [see Eq. (\ref{eqn:fieldtensor})]
\begin{multline}
\Ftilde^{\mu\nu}(\phi,\phi_0)
=
\int^{\phi}_{\phi_0} d\phi'\, F^{\mu\nu}(\phi')
\\=
\sum_{i=1,2} f_i^{\mu\nu} [\psi_i(\phi)-\psi_i(\phi_0)]
\end{multline}
[we also use the notation $\Ftilde^{\mu\nu}(\phi) = \Ftilde^{\mu\nu}(\phi,-\infty)$, see \cite{meuren_polarization_2013}].

For an electron which propagates from $\phi_0$ to $\phi$ we define the dressed momentum by \cite{landau_quantum_1981}
\begin{gather}
Q^\mu(\phi,\phi_0) = \frac{1}{(\phi-\phi_0)} \int_{\phi_0}^\phi d\phi' \, P^\mu(\phi').
\end{gather}
Correspondingly, the square of the dress mass is in general given by \cite{mitter_quantum_1975}
\begin{multline}
\label{eqn:polop_dressedmass}
m_*^2(ky,kx) = Q^2(ky,kx) 
\\= 
m^2 \Big[1 + \sum_{i=1,2} \xi_i^2 \big( J_i-I_i^2 \big) \Big],
\end{multline}
where $I_i$ and $J_i$ are defined in Eq.~(\ref{eqn:polop_massdressing}). As it depends only on $e^2$, the positron has the same dressed mass. For $\psi_1(\phi) = \sin(\phi)$, $\psi_2=0$, $kx=0$, $ky=2\pi$ we obtain the above monochromatic result. 

Thus, the nonlinear phase of the polarization operator [see Eq.~(\ref{eqn:polop_phases})] can be interpreted in terms of the mass dressing in the laser field \cite{becker_vacuum_1975,dinu_vacuum_2014}.

\section{Numerical calculation of oscillatory integrals}
\label{sec:numericalintegration}

The double-integral representation for the polarization operator given in Eq.~(\ref{eqn:polarizationoperator_realincoming_Wfunctions}) contains oscillatory integrals in the variables $\varrho$ and $y^\lminus$. The integral in $y^\lminus$ can be calculated using the fast fourier transform (FFT) \cite{cooley_algorithm_1965,fftw_2005}. The integral in $\varrho$, however, is more complicated. While the phase $\widetilde{\Phi}_0$ oscillates regularly, the phase $\widetilde{\Phi}_1$ is nonlinear [due to the appearance of the field-dependent function $\mc{D}(\varrho,ky)$, see Eq.~(\ref{eqn:polop_doubleint_realincoming_phases})]. However, the derivative
\begin{gather}
\label{eqn:polop_Dprime_ky}
\frac{\del \mc{D}(\varrho,ky)}{\del \varrho}
=
1 + \sum_{i=1,2} \xi_i^2 [\psi_i(ky - 2 \varrho) - I_i(\varrho,ky)]^2
\end{gather}
is always positive \highlight{(this has also been observed in \cite{dinu_vacuum_2014})} and therefore the change of variables $u = \mc{D}(\varrho)$ can be applied to obtain an regularly oscillating integral \cite{evans_alternative_1994}
\begin{multline}
\label{eqn:polop_transformationtofourierint}
\int_0^\infty \, \frac{d\varrho}{\varrho} g(\varrho) e^{-\I 4(\nfrac{m^2}{kq}) \mc{D}(\varrho) }
\\=
\int_0^\infty \, \frac{du}{\mc{D}'(\varrho)} \frac{g(\varrho)}{\varrho} e^{-\I 4 (\nfrac{m^2}{kq}) u },
\end{multline}
where the inverse function $\varrho = \mc{D}^{-1}(u)$ is calculated numerically [due to $\mc{D}'(\varrho) > 0$ the map is one-to-one]. Having applied this change of variables, we obtain an ordinary Fourier integral, which can be evaluated with standard methods.

Fourier integrals (with finite limits) can be calculated very fast using Chebyshev series expansions \cite{piessens_computation_1975,piessens_quadpack_1983,piessens_computation_1984}. To this end we write
\begin{gather}
\label{eqn:fourierintegral_finiteboundaries}
\int_a^b dx \, e^{i\omega x} g(x)
= 
\delta e^{\I \omega c} \int_{-1}^{+1} dt \, e^{\I \Omega t} f(t),
\end{gather}
where we used the change of variables $x(t) = c + \delta t$ with $c = \nfrac{(a+b)}{2}$, $\delta =\nfrac{(b-a)}{2}$ and defined $f(t) = g[x(t)]$, $\Omega = \delta \omega$. If the function $f(t)$ is slowly varying, its expansion into a Chebyshev series is rapidly converging \cite{olver_nist_2010,boyd_chebyshev_2001}
\begin{gather}
f(t) = \sideset{}{'}\sum^{\infty}_{n=0} c_n T_n(t),
\quad
T_{n}(t) = \cos(n\theta),
\quad
t = \cos \theta,
\end{gather}
where
\begin{multline}
c_n 
= 
\frac{2}{\pi}  \int_{-1}^{+1} dt \, \frac{T_n(t) f(t)}{\sqrt{1-t^2}} 
\\= 
\frac{2}{\pi}  \int_{0}^{\pi} d\theta \, \cos(n\theta) f(\cos \theta) 
\end{multline}
(the prime at the sum symbol indicates that the first coefficient in the sum is halved). The Chebyshev series coefficients can be calculated using FFT. The absolute error due to the truncation of the Chebyshev series can be estimated from the last series coefficients \cite{boyd_chebyshev_2001}.

Having computed the series coefficients, the Chebyshev moments
\begin{gather}
\begin{aligned}
C_n(z)
&=
\int_{-1}^{+1} dt \, T_{2n}(t) e^{\I z t},
\\
S_n(z)
&=
\I \int_{-1}^{+1} dt \, T_{2n+1}(t) e^{\I z t}
\end{aligned}
\end{gather}
must be calculated in order to evaluate the integral in Eq.~(\ref{eqn:fourierintegral_finiteboundaries}). To this end we note that they obey the following three-term recurrence relations \cite{piessens_quadpack_1983}
\begin{subequations}
\label{eqn:chebyshevmoments_3termrecurrence2}
\begin{multline}
z^2 (n-1)(2n-1) C_{n+1}(z)
\\- 
(n+1)(n-1) \lsb 4z^2 - 8(2n+1)(2n-1) \rsb C_n(z)
\\+
z^2 (n+1)(2n+1) C_{n-1}(z)
\\=
-16 (n-1)(n+1) \cos(z)
+
12 z \sin(z),
\end{multline}
\begin{multline}
z^2 (2n-1)n S_{n+1}(z)
\\- 
(2n+3)(2n-1) \lsb z^2  - 8n(n+1) \rsb S_n(z)
\\
+z^2 (2n+3)(n+1) S_{n-1}(z)
\\=
4 (2n-1)(2n+3) \sin(z)
+
12 z \cos(z).
\end{multline}
\end{subequations}
For certain parameters (e.g. for very large frequencies) the Chebyshev moments can be calculated by applying the above relations in the forward direction (e.g. $S_n$ can be calculated by starting from $S_0$ and $S_1$). However, this procedure is in general numerically unstable and Olver's algorithm must be used \cite{olver_numerical_1967,wimp_computation_1984}. By calculating $C_n$ and $S_n$ independently, we can estimate the numerical error of the calculated Chebyshev moments by evaluating the relation \cite{piessens_quadpack_1983}
\begin{gather}
S_n(z) = \frac{\sin z}{2(n+1)n} - \frac{z}{4n} C_n(z) + \frac{z}{4(n+1)} C_{n+1}(z).
\end{gather}

\end{document}